\documentclass{aa}  
\usepackage{graphicx}
\usepackage{txfonts}
\begin{document}

   \title{Probing the Milky Way Halo with RR Lyrae Stars from {\it Gaia} Data Release 3}

   \author{T. Muraveva\inst{1}
          \and 
          L. Monti\inst{1}
          \and
          D. Massari\inst{1}
          \and
          M. De Leo\inst{2}
          \and
          A. Garofalo\inst{1}
          \and
          G. Clementini\inst{1}
          \and
          E. Ceccarelli\inst{1, 2}
          \and
          U. Michelucci\inst{3,4}
          }

   \institute{INAF - Osservatorio di Astrofisica e Scienza dello Spazio di Bologna, via Piero Gobetti 93/3, Bologna 40129, Italy\\
   \email{tatiana.muraveva@inaf.it}
   \and
   Dipartimento di Fisica e Astronomia-Universitá di Bologna, via Piero Gobetti 93/2, Bologna 40129, Italy
   \and
   Computer Science Department, Lucerne University of Applied Sciences and Arts, Luzern 6002, Switzerland
   \and
   TOELT LLC, Machine Learning Research and Development Department, Winterthur 8406, Zurich, Switzerland
   }

   \date{Received XXXX; accepted XXXX}

 
   \abstract
{The Milky Way (MW) stellar halo, containing debris from past accretion events, serves as a fossil record of hierarchical mass assembly. Due to their distinct properties, RR Lyrae stars (RRLs) serve as excellent tracers for identifying and characterising the halo's substructures. With the advent of {\it Gaia} Data Release 3 (DR3), which includes high-precision positions, parallaxes, proper motions, radial velocities, the identification and characterisation of thousands of RRLs, it has become possible to study the distribution, kinematics, and metallicity of RRLs in the various dynamical substructures with unprecedented detail.}
{Our primary goal is to identify and characterise the dynamical substructures of the MW halo using RRLs from {\it Gaia} DR3.}
{We analysed a sample of 4933 RRLs, for which we calculated the integrals of motion and orbital parameters. We applied the domain-informed novelty detection CLustering in Multiphase Boundaries
 (CLiMB) framework to identify RRL membership in the MW substructures. We then used newly calibrated photometric metallicities available in the literature to study the metallicity distributions of RRLs in different substructures.}
{We analysed the metallicity distributions of RRLs in major accreted system remnants as a snapshot of their chemical evolutionary status during early epochs. We calculated the weighted mean metallicity ([Fe/H]) and the corresponding standard deviation for {\it Gaia} Sausage/Enceladus ([Fe/H] = $-1.57 \pm 0.25$ dex), Sequoia ([Fe/H] = $-1.64 \pm 0.26$ dex), and the Helmi streams ([Fe/H] = $-1.66 \pm 0.19$ dex). The metallicity distribution of RRLs in Thamnos was found to be bimodal, with the metal-poor peak likely representing the genuine accreted Thamnos population ([Fe/H] = $-1.94\pm0.20$ dex), in agreement with recent works based on spectroscopic abundances. 
Our analysis shows that the substructures ED-1 and L-RL3 are highly contaminated by thick disc stars. However, the metal-poor tails in their metallicity distributions may be signatures of remnants from small accreted systems. We also identify over-densities of RRLs in correspondence with the recently reported substructures Shiva and Shakti, which we suggest are of {\it in-situ} origin. Finally, we applied the RRL-based mass–metallicity relation of galaxies to test the nature of the identified dynamical substructures.}
{}


   \maketitle
%

\section{Introduction}

The hierarchical mass assembly history of the Milky Way (MW) is a fundamental prediction of the $\Lambda$ Cold Dark Matter ($\Lambda$CDM) cosmological model, which posits that galaxies grow through successive mergers with smaller systems \citep{White1991}. Consequently, the MW’s stellar halo is expected to be largely composed of debris from accreted dwarf galaxies and globular clusters (GCs), preserving a fossil record of these past interactions. Stars originating from the same disrupted progenitor are predicted to retain coherence in their integrals of motion (IoM) and to display characteristic chemical abundance patterns, even long after phase mixing has occurred \citep{Johnston1996, Helmi2000, Gomez2013, Helmi2020}. Identifying such stellar substructures in the Galactic halo provides powerful constraints on the formation and evolutionary history of the MW.

In recent years, numerous studies have identified substructures in the stellar halo that are believed to be remnants of past accretion events \citep[e.g.][]{Helmi1999, Koppelman2019a, Massari2019, Belokurov2018, Helmi2018, Myeong2019, Kruijssen2020, Naidu2020, Horta2021, Oria2022, Ruiz-Lara2022, Tenachi2022, Malhan2024, Dodd2025a, Dodd2025b}. However, interpreting these discoveries remains challenging. Numerical simulations show that debris from a single merger can generate multiple over-densities in different regions of IoM space, which may be mistakenly interpreted as distinct substructures \citep{Koppelman2020, Amarante2022, Belokurov2023, Davies2023}. Conversely, remnants of different progenitors can significantly overlap in IoM space \citep{Helmi2000, Naidu2020, Lovdal2022, Dodd2023}, while contamination from unrelated sources can strongly affect kinematically selected samples of candidate members \citep{Buder2022, Rey2023, Dodd2025b, Thomas2025}. A promising approach to resolving these issues is to incorporate chemical abundance information, such as the metallicity ([Fe/H]) of stars associated with the substructures \citep{Monty2020, Aguado2021, matsuno22, Horta2023, Ceccarelli2024, Mori2024}.

RR Lyrae stars (RRLs) are pulsating variable stars that serve as highly effective tracers of structures associated with past accretion events of the MW halo due to their distinct properties. They are numerous in the MW halo and easily recognisable thanks to their characteristic light curves and short pulsation periods (typically less than one day). Although recent studies suggest that relatively young and metal-rich RRLs can form through the evolution of close binary systems \citep{Sarbadhicary2021, Bobrick2024}, the majority of RRLs are old (age$\ge$10~Gyr), and thus can provide information on the properties of the various stellar systems before they were accreted by the MW. Moreover, RRLs are excellent distance indicators \citep[e.g.][]{Catelan2004}; and their metallicity can be estimated from photometric parameters, such as the pulsation period and Fourier decomposition parameters of their light curves, without requiring spectroscopic data (e.g., \citealt{Jurcsik1996}; \citealt{Morgan2007}). With the advent of \textit{Gaia} Early Data Release 3 (EDR3; \citealt{Brown2021}), and Data Release 3 (DR3; \citealt{Vallenari2023}), which include high-precision measurements of positions, parallaxes, and proper motions for almost two billion stars, radial velocities for 33 million objects, along with the identification and characterisation of thousands of RRLs \citep{Clementini2023}, it has become possible to study the distribution, kinematics, and metallicity of RRLs in the various dynamical substructures with unprecedented detail.

Numerous studies have used RRLs to investigate different substructures in the MW halo associated with past accretion events. \citet{Cabrera2024} analysed a sample of RRLs with photometric metallicities and distance estimates from \citet{Li2023}, combined with proper motions and radial velocities from {\it Gaia} DR3 \citep{Vallenari2023}. They applied the HDBSCAN clustering algorithm to the IoM, identifying 97 substructures and exploring their associations with known components of the MW, such as {\it Gaia} Sausage/Enceladus (GSE; \citealt{Belokurov2018, Helmi2018}), the metal-weak thick disk, the Helmi streams \citep{Helmi1999}, Sequoia \citep{Myeong2019, Matsuno2019}, Sagittarius stream \citep{Ibata1994}, and Wukong \citep{Naidu2020}. \citet{Kunder2024} analysed a sample of 80 RRLs from the BRAVA-RR survey \citep{Kunder2016, Kunder2020} and attributed a small fraction of them to the accreted GSE population, while the dominant fraction was associated with the {\it in-situ}, metal-poor Galactic component Aurora \citep{Belokurov2022}, which formed before the Galaxy developed a coherently rotating disk. However, \citet{Kunder2024} did not rule out the possibility that a fraction of these stars originated from an ancient accretion event such as Kraken/Heracles \citep{Massari2019, Kruijssen2020, Horta2021}. \citet{Dorazi2024} derived metallicities from high-resolution (HR) spectra for 78 RRLs observed by the GALactic Archaeology with HERMES (GALAH) survey \citep{DeSilva2015, Buder2021} and complemented this dataset with additional RRLs for which HR or low-resolution (LR) spectroscopic metallicities are available in the literature, resulting in an extended catalogue of 535 RRLs. These authors provided preliminary associations of the analysed RRLs with GSE, the Helmi streams, Sequoia, Sagittarius, and Thamnos \citep{Koppelman2019a}, and investigated the metallicity distributions of RRLs in these substructures. Most recently, \citet{Sun2025} identified Galactic substructures in 5D space using RRLs from {\it Gaia} DR3 \citep{Vallenari2023}. They combined positions and proper motions from {\it Gaia} DR3 with photometric metallicities and distances from \citet{Li2023}, and identified several substructures, including GSE, the Helmi streams, Sequoia, and Wukong.

In this paper, we analyse different dynamical substructures of the MW using IoM and photometric metallicities of RRLs from the {\it Gaia} DR3 catalogue, with three key differences compared to previous studies: (1) we adopt the new photometric metallicities from \citet{Muraveva2025}, derived from period–Fourier parameters–metallicity relations calibrated by means of Machine Learning algorithms, using RRLs with accurately measured spectroscopic metallicities \citep{Crestani2021, Gilligan2021, Liu2020}; (2) to identify substructures, we apply a newly developed domain-informed novelty detection CLustering in Multiphase Boundaries
 (CLiMB, Monti et al., in prep.) framework, designed for datasets containing both labelled and unlabelled components. This algorithm is particularly well-suited for RRL studies, as purely unsupervised methods often struggle to perform reliably with sparse, low-density samples such as RRLs; (3) finally, for the first time, we apply the newly derived mass–metallicity relation (MZR) for galaxies based on RRLs \citep{Bellazzini2025} to test the nature of the identified dynamical substructures.

The paper is organised as follows. In Section~\ref{sec:data}, we describe the dataset used in this study. Section~\ref{sec:method} outlines the clustering algorithm utilized to identify the dynamical substructures of the MW. In Section~\ref{sec:metallicity}, we present the metallicity distribution of RRLs in the identified substructures, while Section~\ref{sec:mzr} analyses their MZR. Finally, Section~\ref{sec:concl} provides the conclusions and final remarks.

\section{Dataset}\label{sec:data}

The {\it Gaia} DR3 includes a clean catalogue of 270,891 RRLs analysed by the Specific Object Study (SOS) pipeline for Cepheids and RRLs (SOS Cep\&RRL; \citealt{Clementini2023}). This catalogue provides, among other parameters, periods, amplitudes, mean magnitudes in the $G$, $G_{BP}$, and $G_{RP}$ bands, as well as Fourier decomposition parameters of the $G$-band light curves. In \citet{Muraveva2025}, we further cleaned the {\it Gaia} DR3 sample of RRLs by comparing it with the OGLE IV catalogues \citep{OGLE1, OGLE2} and by analysing the distribution of RRLs on the Bailey diagram ($G$-band amplitude versus pulsation period). This procedure yielded a sample of 258,696 stars, which we consider bona fide RRLs. We then calibrated new empirical relations between the metallicity of RRLs, their pulsation periods, and Fourier decomposition parameters using Machine Learning methods, based on accurate spectroscopic metallicities from the literature \citep{Crestani2021, Gilligan2021, Liu2020}. As a result, we obtained photometric metallicity estimates for 134,769 RRLs from the cleaned {\it Gaia} DR3 sample \citep{Muraveva2025}.

For our analysis, we selected RRLs from the cleaned sample for which coordinates, radial velocities, and proper motions were available in the {\it Gaia} EDR3 \citep{Brown2021} and DR3 catalogues \citep{Vallenari2023}, while geometric distances were estimated from the {\it Gaia} EDR3 parallaxes using a Bayesian approach \citep{BJ2021}. This selection yielded a sample of 4933 RRLs, which we used as the reference catalogue (Table~\ref{tab:gen}) in this study. We chose parallax-based distances instead of distances derived from the RRL fundamental relations, such as the luminosity-metallicity relation in the visual band (e.g., \citealt{Clementini2003}; \citealt{Bono2003}) or the near/mid-infrared period-luminosity-metallicity relations (e.g., \citealt{Longmore1986}; \citealt{Sollima2008}; \citealt{Muraveva2018}), for our sample of 4933 RRLs to avoid introducing potential systematic biases from metallicity-dependent fundamental relations into the measurement of distances. This approach ensures that the method with which we assign RRLs to different dynamical substructures is independent of the metallicity of the individual RRLs, allowing for a robust analysis of metallicity distributions across the identified substructures.
Additionally, since our sample is primarily limited by the availability of radial velocity measurements rather than distance constraints, and the vast majority of RRLs in the reference sample have accurate parallax data (uncertainties <0.1 mas for 4925 stars out of 4933), measuring distances from parallaxes provides a reliable and straightforward option.

We calculated the Galactocentric Cartesian coordinates ($X$, $Y$, $Z$) for the RRLs in our sample using their positions and estimated distances. The Sun was assumed to lie on the $X$-axis of a right-handed coordinate system. In this system, the $X$-axis points from the Sun toward the Galactic centre, the $Y$-axis points toward Galactic longitude $l = 90^{\circ}$, and the $Z$-axis points toward the North Galactic Pole ($b = 90^{\circ}$). The distance of the Sun from the Galactic centre was assumed to be 8.122~kpc \citep{Gravity2018}. 

We then computed the IoM, including the total energy ($E$), the angular momentum along the $Z$-axis ($L_z$), and the component of angular momentum perpendicular to $L_z$ ($L_\perp$), as well as orbital parameters (e.g., apocentre, pericentre, and eccentricity) for 4933 RRLs in the sample. These computations were performed using the AGAMA software package \citep{AGAMA} and adopting the MW potential from \citet{McMillan2017}. For each RRL, we ran 100 Monte Carlo simulations of the orbit, assuming Gaussian uncertainties in distance, proper motion, and radial velocity. The final values of the dynamical parameters were derived as the median of their distributions. In the following analysis, we use the IoM, Cartesian coordinates, and photometric metallicities of RRLs in the reference catalogue (Table~\ref{tab:gen}) to study the dynamical substructures of the MW.

\section{Clustering analysis}\label{sec:method}

\begin{figure*}
\includegraphics[width=18cm]{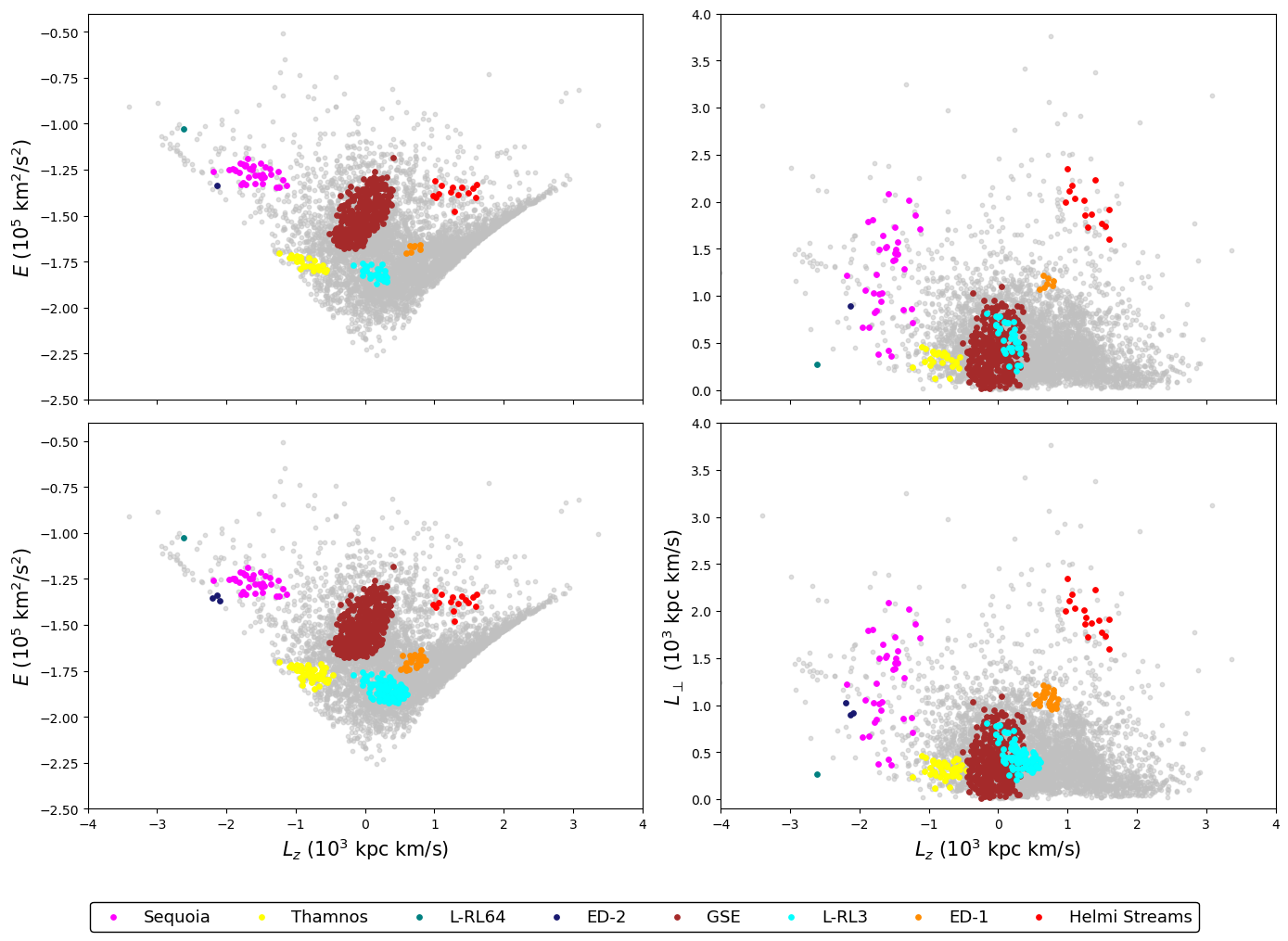}
\caption{Distribution of 4933 RRLs from our sample in the $E$-$L_z$ and $L_z$-$L_\perp$ planes, colour-coded by the substructure to which they belong. Grey dots represent RRLs not assigned to any substructure. The top panels show RRLs identified in known substructures by cross-matching with D23, while the bottom panels display RRLs assigned to known substructures during  the first phase of the CLiMB algorithm.}\label{fig:clusters_all_old}
\end{figure*}

\begin{figure*}
\includegraphics[width=18cm]{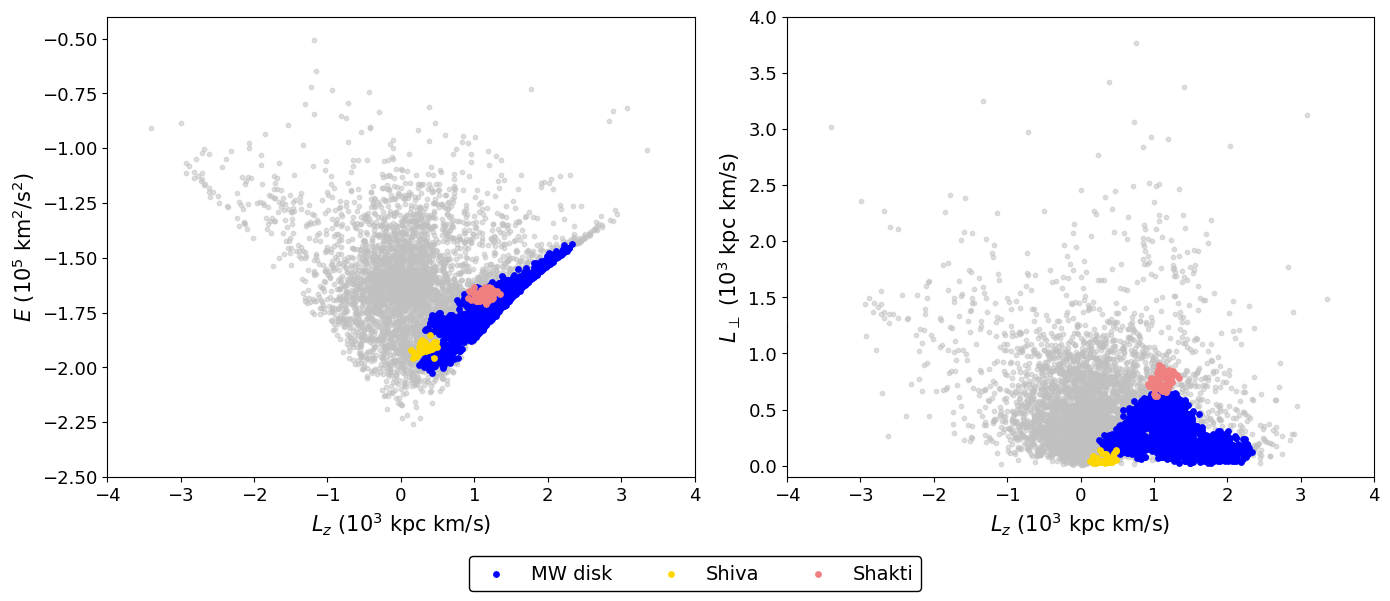}
\caption{Distribution of 4933 RRLs from our sample in the $E$–$L_z$ and $L_z$–$L_\perp$ planes, colour-coded by the substructure to which they were assigned during the second phase of the CLiMB algorithm. Grey dots represent RRLs not assigned to any substructure}\label{fig:clusters_all_new}
\end{figure*}

Prior to the execution of the clustering analysis, we implemented a specific preprocessing routine to mitigate biases and ensure isotropic metric properties. First, to mitigate systematic biases related to the absolute scale of the $E$ observable, that can introduce artificial correlations, we centered the total energy distribution. 
Given the vector of $E$ values across all RRLs ($e$), the mean background total energy level $\mu_e$ was subtracted from the original values ($e' = e - \mu_e$). This procedure effectively removes the mean background energy level while preserving local kinematic structures. Subsequently, given the heterogeneous ranges of the features ($E$, $L_z$, $L_\perp$), the dataset was standardized using Z-score normalization (zero mean, unit variance) via \texttt{StandardScaler} from scikit-learn\footnote{\url{https://scikit-learn.org}} to ensure isotropic distance calculations.

To investigate the dynamical structures in the MW traced by RRLs, we applied the CLiMB library\footnote{\url{https://github.com/LorenzoMonti/CLiMB}} (Monti et al., in prep.), which implements a novel domain-informed novelty detection clustering algorithm. CLiMB is a semi-supervised, two-phase clustering algorithm designed for datasets containing both labelled and unlabelled components. In the first phase (K-bound), the algorithm incorporates prior knowledge by considering RRLs previously associated with known MW substructures, as reported in the literature, and identifies additional RRLs in the sample that are likely members of these structures. It performs constrained clustering using an adapted K-means approach, entirely redesigned with the inclusion of boundaries and utilizing the Mahalanobis distance metric, which is a multivariate distance measure that considers the covariance among variables. 
The Mahalanobis metric is essential when identifying stellar substructures in the space of IoM. Unlike the Euclidean metric, which assumes spherical clusters, it properly accounts for correlations and differing scales between features. The accreted structures typically appear as elongated debris rather than globular distributions. By incorporating the covariance matrix, CLiMB defines ellipsoidal boundaries that follow the natural dynamical manifold of stellar streams.


CLiMB incorporates prior knowledge through initial centroid positions (e.g., from literature-identified RRLs associated with known MW substructures) and enforces strict density, distance, and radial constraints. To ensure reproducibility and robustness, the hyperparameters governing these constraints were optimized by maximizing the Adjusted Rand Index (ARI), a clustering similarity metric that quantifies the agreement between the algorithm output and a reference classification while correcting for chance, calculated on the subset of known substructures. The density constraint filters out sparse points by requiring a minimum local density (\texttt{density\_threshold} $= 0.005$), ensuring that only points within well-defined regions are considered for assignment. The distance constraint requires that points lie within a maximum radius of their nearest centroid (\texttt{distance\_threshold} $= 0.5$). Finally, the radial constraint prevents centroids from drifting significantly from their initial, knowledge-guided positions (\texttt{radial\_threshold} $= 0.1$ and \texttt{convergence tolerance $= 0.01$}), ensuring robust convergence and adherence to prior domain information. 
 
RRLs not assigned during K-bound (i.e., those failing any of the aforementioned constraints) are passed to the second phase, so-called Exploratory Clustering. Here, density-based methods (such as DBSCAN) are applied to discover new, potentially irregular clusters among these previously unassigned stars. Using the optimal parameters from the first stage, we optimized the exploratory parameters by minimizing the Davies-Bouldin Index (DBI), an internal clustering validation metric that measures the average separation between clusters relative to their internal dispersion \citep{davies2009cluster} on the clusters discovered within the unclassified data. This yielded an optimal configuration of $\epsilon=0.190$ and $min\_samples=24$. This allows CLiMB to identify previously unknown substructures within the remaining RRL sample, thereby combining the benefits of guided analysis with unsupervised discovery of additional stellar clusters.
 
Recently, Dodd et al. (\citeyear{Dodd2023}, hereafter D23) analysed a sample of 193,831 nearby MW halo stars for which accurate parallaxes and radial velocities are available in {\it Gaia} DR3. Using a single-linkage clustering algorithm in IoM space ($E$, $L_z$, $L_\perp$), together with Mahalanobis distances and metallicities, D23 identified seven major groups and eleven smaller independent clusters, many of which were already known from the literature. We cross-matched our catalogue of RRLs, compiled as described in Section~\ref{sec:data}, with the sample of D23 and found 2009 stars in common. Among these, we selected 468 RRLs that, according to D23, belong to one of the eight large known groups or clusters: GSE, L-RL3, Thamnos, Helmi streams, Sequoia, ED-1, L-RL64, and ED-2. The second column of Table~\ref{tab:clusters_dodd} lists the number of RRLs in each substructure as classified by D23, while the top panels of Figure~\ref{fig:clusters_all_old} show their distribution in the $E$–$L_z$ and $L_z$–$L_\perp$ planes.

In the first phase of the CLiMB algorithm, all 468 RRLs are assigned the substructure labels of D23. These labelled stars serve as the initial training set for a semi-supervised clustering procedure, which then searches for additional members of the same substructures within the reference sample based on their distribution in IoM ($E$, $L_z$, $L_\perp$) space. Table~\ref{tab:clusters_dodd} lists the number of stars in each substructure found by CLiMB, along with the percentage increase relative to the number of RRLs identified in each substructure by D23. The percentage increase ranges from 0\% (no new RRLs identified) for L-RL64 and Sequoia, to 286\% (the number of RRLs in the substructure increased by nearly four times) for ED-1. In total, 630 RRLs, shown in the bottom panels of Fig.~\ref{fig:clusters_all_old}, were assigned to known substructures by the CLiMB algorithm, 35\% more than the 468 RRLs initially identified by D23.

In the second exploratory phase of the CLiMB algorithm, we removed the 630 RRLs identified as members of known substructures from the reference sample (Table~\ref{tab:gen}) and applied the DBSCAN algorithm with the optimized parameters described above to the remaining stars to search for new structures not reported by D23. This phase yielded three additional groups, shown in Table~\ref{tab:clusters_new} and Fig.~\ref{fig:clusters_all_new}. The largest substructure, shown in blue, occupies the region corresponding to the MW disk. 
The ability of CLiMB to recover the disk as an individual cluster supports the robustness of our method. However, the disk structure identified by CLiMB likely includes both the thin disk and the thick disk, as discussed further in Section~\ref{subsec:disk}. During the exploratory phase, we also identified two additional clusters, marked in pink and yellow in Fig.~\ref{fig:clusters_all_new}. In Section~\ref{subsec:shiva_shakti}, we show that these clusters correspond to two newly reported substructures, Shiva and Shakti \citep{Malhan2024}. 

In total, our semi-supervised, two-phase CLiMB algorithm enabled the identification of 1866 RRLs in 11 MW substructures, which are listed in Tables~\ref{tab:clusters_dodd} and \ref{tab:clusters_new}. The remaining 3067 RRLs from the reference sample were not assigned to any substructure. As a test, we remove RRLs identified in the 11 substructures and re-run the exploratory phase of the CLiMB algorithm on the remaining sample of RRLs. No new clusters were identified.
In the following section, we analyse in more detail the metallicity distribution of RRLs in the 11 substructures.

\begin{table}
\caption{RRLs in different MW substructures identified during the first phase of the CLiMB algorithm.}
\label{tab:clusters_dodd}
\begin{tabular}{lccc}
\hline
Substructure & $N$ & $N$ & $\Delta  N $ \\
 & (D23) & (CLiMB) & (\%) \\
\hline
ED-1 & 7 & 27 & 286\% \\
ED-2 & 1 & 3 & 200\% \\
GSE & 350 & 398 & 14\%  \\
Helmi streams & 14 & 16 & 14\%  \\    
L-RL3 & 34 & 102 & 200\% \\
L-RL64 & 1 & 1 & 0\% \\
Sequoia & 35 & 35 & 0\% \\
Thamnos & 26 & 48 & 85\%\\
\hline
Total & 468 & 630 & 35\% \\
\hline
\end{tabular}
\tablefoot{Column (1): Name of the substructure.
Column (2): Number of RRLs identified in the substructure by cross-matching with the D23 catalogue.
Column (3): Number of RRLs identified in the substructure using the CLiMB algorithm.
Column (4): Percentage increase in the number of RRLs identified in each substructure by the CLiMB algorithm compared to D23.}
\end{table}

\begin{table}
\caption{RRLs in different MW substructures identified during the second phase of the CLiMB algorithm.}
\label{tab:clusters_new}
\centering
\begin{tabular}{lc}
\hline
Substructure & $N$ \\
\hline
Disk CLiMB & 1140 \\
Shiva & 38 \\
Shakti & 58 \\
\hline
Total & 1236 \\
\hline
\end{tabular}
\tablefoot{Column (1): Name of the substructure.
Column (2): Number of RRLs identified in the substructure using the CLiMB algorithm.}
\end{table}


\section{Metallicity distribution}\label{sec:metallicity}

\begin{table*}
\caption{Mean metallicities of RRLs in different substructures of the MW}\label{tab:clusters_metallicity}
\begin{tiny}
\begin{tabular}{l c c c c c c c}

\hline
Substructure & $N$ & $[\mathrm{Fe/H}]$ & $N$ & $[\mathrm{Fe/H}]$ & $[\mathrm{Fe/H}]$  &  $[\mathrm{Fe/H}]$ &  $[\mathrm{Fe/H}]$ \\
 & (CLiMB) & (CLiMB)  & (D23) & (D23)  & & & \\
 &  & this study &  & this study & \citep{Dorazi2024} & \citep{Cabrera2024}& \citep{Sun2025} \\
\hline

Sequoia & 29 & $-1.64 \pm 0.26$ & 29 & $-1.64 \pm 0.26$ & $-1.60\pm0.33$ & $-1.70\pm 0.26$ & $-1.72\pm0.32$ \\
Thamnos: & 33 & $-1.76 \pm 0.33$ & 22 & $-1.75 \pm 0.36$ & $-1.75 \pm 0.28$\tablefootmark{(*)} & " " & " " \\
- metal-rich component & 10 & $-1.35\pm0.12$ & " " & " "& " "  & " " & " " \\
- metal-poor component & 23 & $-1.94\pm0.20$ & " " & " "& " "  & " " & " " \\
L-RL3 & 71 & $-1.54 \pm 0.33$ & 25 & $-1.56 \pm 0.38$ & " " &" " & " "\\
L-RL64 & 1 & $-1.57 \pm 0.41$ & 1 & $-1.57 \pm 0.41$ & " " & " "& " " \\
ED-2 & 2 & $-1.94 \pm 0.32$ & 1 & $-2.22 \pm 0.34$ & " " & " " & " " \\
GSE & 315 & $-1.57 \pm 0.25$ & 273 & $-1.56 \pm 0.25$ & $-1.65\pm0.29$ & $-1.61\pm0.31$ & $-1.62 \pm 0.32$ \\
ED-1 & 18 & $-1.57 \pm 0.40$ & 6 & $-1.42 \pm 0.28$ & " " & " " & " " \\
Helmi streams & 12 & $-1.66 \pm 0.19$ & 10 & $-1.66 \pm 0.19$ & $-1.67\pm0.29$& $-1.78\pm 0.33$ & $-1.78 \pm 0.30$\\
\hline
Disk CLiMB:& 838 & $-1.12 \pm 0.48$ & " " & " " & " " & " "  & " " \\
 - thin disk & 238 & $-0.75 \pm 0.40$ & " " & " " & " " &  " " & " " \\
 - thick disk & 600 & $-1.27 \pm 0.42$ & " " & " " & " " & " " & " " \\
Shiva & 30 & $-1.38 \pm 0.29$ & " " & " " & " " & " " & " " \\
Shakti & 40 & $-1.41 \pm 0.38$ & " " & " " & " " & " " & " " \\
\hline
\end{tabular}
\end{tiny}
\tablefoot{Column (1): Name of the substructure.
Column (2): Number of RRLs in each substructure identified with the CLiMB algorithm, for which photometric metallicities from \citet{Muraveva2025} are available and have uncertainties smaller than 0.5~dex.
Column (3): Weighted mean metallicity of RRLs identified as belonging to the substructure using the CLiMB algorithm, with uncertainties calculated as the weighted standard deviation of the mean. For substructures containing only one star, the uncertainty corresponds to the photometric metallicity uncertainty of that individual star.
Column (4): Same as Column (2), but for RRLs identified as part of the substructure through cross-matching with the D23 catalogue.
Column (5): Same as Column (3), but for RRLs identified as part of the substructure through cross-matching with the D23 catalogue.
Column (6): Mean metallicity of RRLs from \citet{Dorazi2024}.
Column (7): Mean metallicity of RRLs from \citet{Cabrera2024}.
Column (8): Mean metallicity of RRLs from \citet{Sun2025}.\\
\tablefootmark{(*)} \citet{Dorazi2024} provided the mean metallicities for Thamnos I and Thamnos II separately. In the table, we report the average of the two values.}
\end{table*}

\subsection{Comparison with the literature}\label{subsec:comp}

\begin{figure*}
\includegraphics[width=18cm]{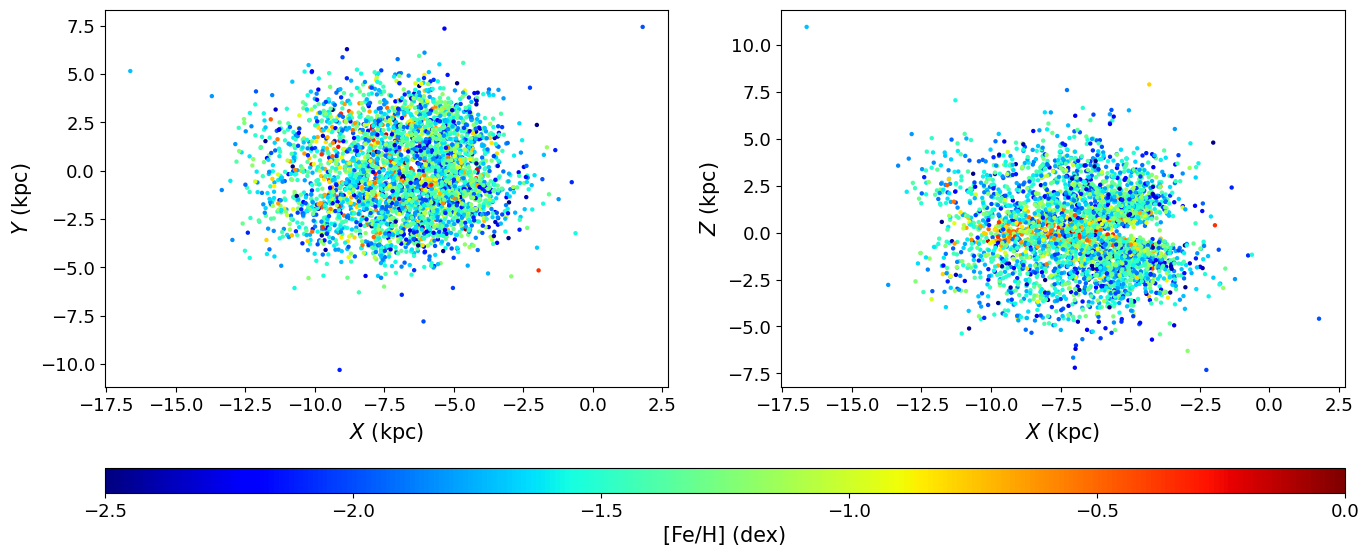}
\caption{Distribution of 3614 RRLs from the reference sample, for which uncertainties in photometric metallicities are less than $0.5~\text{dex}$, on the Cartesian $Y$- $X$ ($\textit{left}$) and $Z$-$X$ ($\textit{right}$) planes, colour-coded by metallicity.} \label{fig:xyz}
\end{figure*}

Metallicity values from \citet{Muraveva2025} are available for 4100 RRLs out of the 4933 in our sample. From these, we selected RRLs with metallicity uncertainties smaller than 0.5 dex, resulting in a final sample of 3614 RRLs, with a weighted mean metallicity and corresponding standard deviation of $-1.47\pm0.44$~dex. Fig.~\ref{fig:xyz} shows their distribution on the Cartesian $Y$-$X$ and $Z$-$X$ planes, colour-coded by metallicity.The third column of Table~\ref{tab:clusters_metallicity} reports the mean metallicity of RRLs assigned to each substructure identified in Section~\ref{sec:method} using the CLiMB algorithm.
As a consistency check, the fifth column provides the mean metallicities of RRLs assigned to each substructure solely through cross-matching with the D23 catalogue, excluding those added by CLiMB (Section~\ref{sec:method}).
As can be seen, the metallicity values are generally in very good agreement (within 0.02~dex) for all but two substructures, confirming that the inclusion of new stars via CLiMB does not significantly affect the metallicity distribution of RRLs. This supports the robustness of our method. However, for two substructures (ED-2 and ED-1), the mean metallicity estimates differ by more than 0.15~dex, although they remain consistent within the uncertainties. We discuss these differences in more detail in Sections~\ref{subsec:ed1_lrl3} and \ref{subsec:ed2}.

In Table~\ref{tab:clusters_metallicity}, we also show the mean metallicity of RRLs in different substructures as reported in the literature \citep{Dorazi2024, Cabrera2024, Sun2025}. 
We find that our mean metallicity values are in good agreement (within 0.1~dex) with the HR and LR spectroscopic metallicities reported by \citet{Dorazi2024}.
In contrast, comparison of our estimates with the mean metallicities from \citet{Cabrera2024} and \citet{Sun2025} shows that our values are slightly more metal-rich, though still consistent within the uncertainties. Both of these studies adopt photometric metallicities from \citet{Li2023}. As discussed in \citet{Muraveva2025} (see section~4.1), the individual metallicities of first-overtone RRLs from \citet{Li2023} are systematically lower than those from \citet{Muraveva2025}, most likely due to differences in the training sets used to calibrate the period–Fourier parameters–metallicity relations in the two studies. Therefore, the offset we observe between our mean metallicities and those reported by \citet{Cabrera2024} and \citet{Sun2025} can be attributed to the difference in the adopted metallicity scales. In the following sections, we analyse the metallicity distributions in the different substructures in more detail.

\subsection{MW disk}\label{subsec:disk}

\begin{figure*}
\includegraphics[width=18cm]{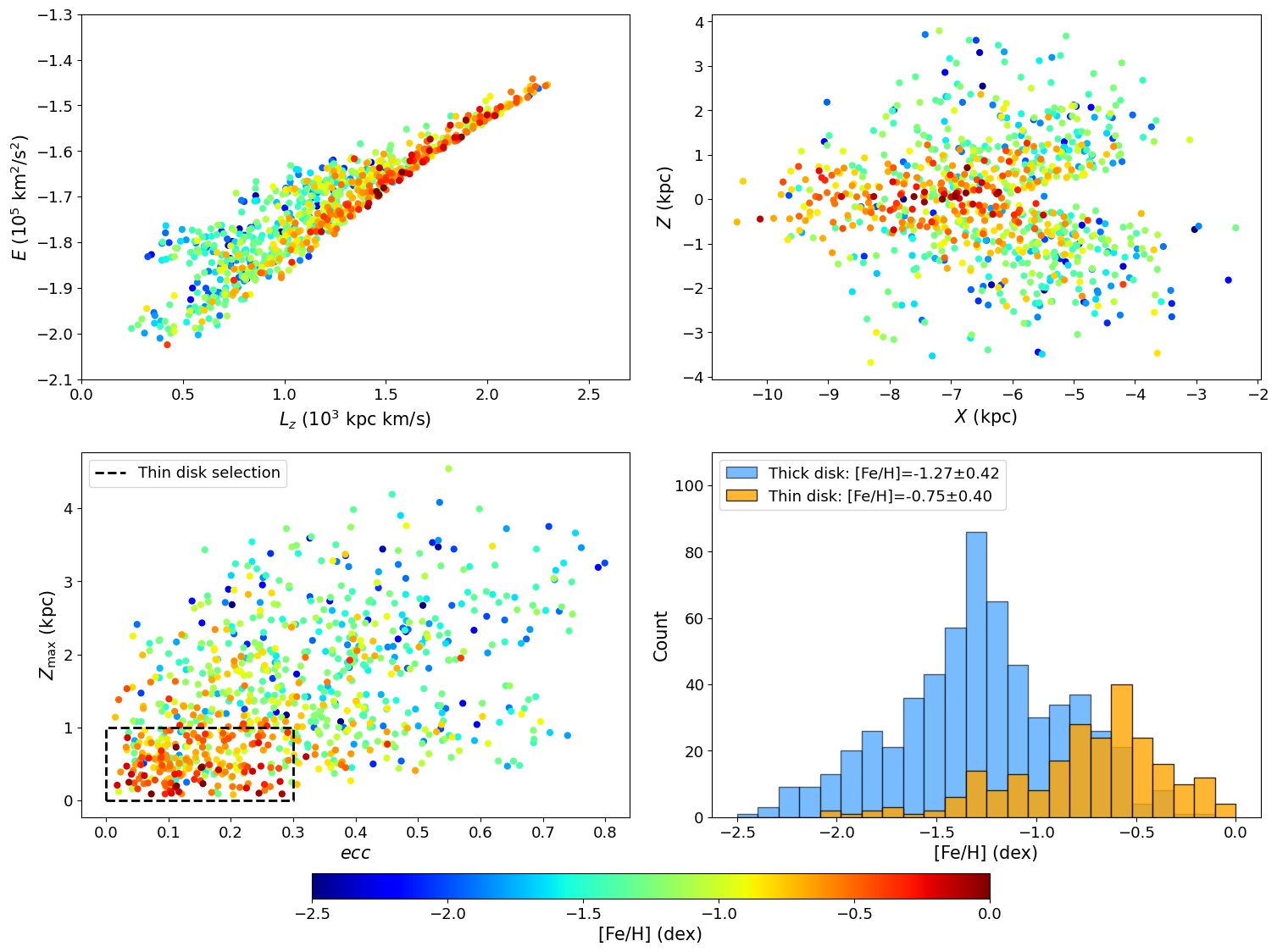}
\caption{Distribution of RRLs in the disk-like structure identified with the CLiMB algorithm  in the $E$ versus $L_z$ ({\it upper left}), Cartesian $Z$ versus $X$ ({\it upper right}), and $Z_{\mathrm{max}}$ versus eccentricity ({\it bottom left}) planes, colour-coded by metallicity. The black dashed line outlines the region used to select the thin-disk RRLs. {\it Bottom right:} Metallicity distributions of RRLs in the thin and thick disks, selected based on their $Z_{\mathrm{max}}$ and eccentricity. See text for details.} \label{fig:disk_metallicity}
\end{figure*}

The thin disk is the dynamically cold component of the MW and the site of ongoing star formation \citep{Helmi2020}. It was initially thought that the thin disk had been forming stars for 8–9 Gyr \citep{Tononi2019}. However, recent studies suggest that the thin disk may have formed less than 1 billion years after the Big Bang and has grown continuously since then, pushing back its onset by about 4–5 billion years compared to earlier estimates of 8–9 Gyr \citep{Nepal2024}.
Kinematically, thin-disk stars are characterized by low eccentricities ($ecc$) and remain confined close to the Galactic plane, with a maximum vertical distance a star reaches above or below the Galactic plane during its orbit ($Z_\mathrm{max}$) peaking at $\sim$0.36 kpc \citep{Deepak2024}.

In contrast, the thick disk is a dynamically hotter and more vertically extended population \citep{Helmi2020}, with a $Z_\mathrm{max}$ distribution peaking at $\sim$1.52 kpc \citep{Deepak2024}. 
Analysis based on {\it Gaia} Data Release 2 (DR2, \citealt{Brown2018}) supports the interpretation that the thick disk primarily formed during or after the merger with GSE \citep{Gallart2019}.  Stars in the thick disk exhibit moderate eccentricities, peaking around 0.42, with most having $ecc< 0.8$ \citep{Deepak2024}. The hot thick disk (HD), sometimes referred to as the Splash population \citep{DiMatteo2019, Belokurov2020}, consists of {\it in-situ} disk stars dynamically heated by the GSE merger. HD stars exhibit hotter kinematics than the thick disk, with median eccentricities of $ecc \approx 0.35$ and a distribution notably skewed toward high-eccentricity stars, including a subset with $ecc = 0.5-1.0$ \citep{Deepak2024}. Most HD stars are located within $Z_\mathrm{max} < 6$ kpc, and the population has a median $Z_\mathrm{max}$ of 2.58 kpc \citep{Deepak2024}.

It is noteworthy to mention that the picture provided to describe all these {\it in-situ} components has recently encountered alternative scenarios. For example, it has been shown that clumpy star formation in the MW proto-disk can cause stars to be dynamically heated via clump instabilities, and to develop into a chemo/kinematically distinct thick disc (which would be coeval with the thin disc, see \citealt{Beraldo2021}). Such instabilities might also explain the emerging of a Splash-like population,  without the need for a significant merger \citep{Amarante2020}.

During the exploratory phase of the CLiMB algorithm (Section~\ref{sec:method}), we identified an extended disk-like structure (blue dots in Fig.~\ref{fig:clusters_all_new}). It likely includes stars from both the thin and thick disks, and possibly a small fraction of the HD. Fig.~\ref{fig:disk_metallicity} shows the distribution of RRLs in this disk-like structure, as identified by CLiMB, in the $E$ versus $L_z$, Cartesian $Z$ versus $X$, and $Z_{\mathrm{max}}$ versus $ecc$ planes, colour-coded by metallicity. As expected, dynamically cold, more metal-rich RRLs have lower eccentricities and are concentrated near the Galactic plane ($|Z| < 1$~kpc), corresponding to the thin disk. In contrast, dynamically warm, more metal-poor RRLs exhibit a more extended spatial distribution ($|Z| < 4$~kpc) and have higher eccentricities, corresponding to the thick disk and likely including a small fraction of the HD.

Although a clear separation remains challenging, we made a tentative attempt to distinguish between the thin- and thick-disk RRL populations based on their orbital eccentricity and $Z_{\mathrm{max}}$. We adopt the criterion that RRLs with $ecc < 0.3$ and $Z_{\mathrm{max}} < 1$~kpc belong to the thin disk. The selected region is outlined by black dashed lines in the bottom-left panel of Fig.~\ref{fig:disk_metallicity}. RRLs located outside this region are likely members of the thick disk and HD. As we are unable to clearly disentangle the thick-disk and HD RRLs in our sample, we refer to this combined component simply as the “thick disk” in the following analysis. Accordingly, we divide the disk-like structure identified with CLiMB into two components: the thin disk, comprising 238 RRLs with a mean metallicity of [Fe/H] = $-0.75 \pm 0.40$~dex, and the thick disk, comprising 600 RRLs with a mean metallicity of [Fe/H] = $-1.27 \pm 0.42$~dex. The bottom-right panel of Fig.~\ref{fig:disk_metallicity} presents the metallicity distributions of RRLs in the thin and thick disks. As expected, the metallicity distribution of the thin-disk RRLs is notably skewed toward higher [Fe/H] values.

\subsection{GSE}\label{subsec:GSE}

\begin{figure*}
\includegraphics[width=18cm]{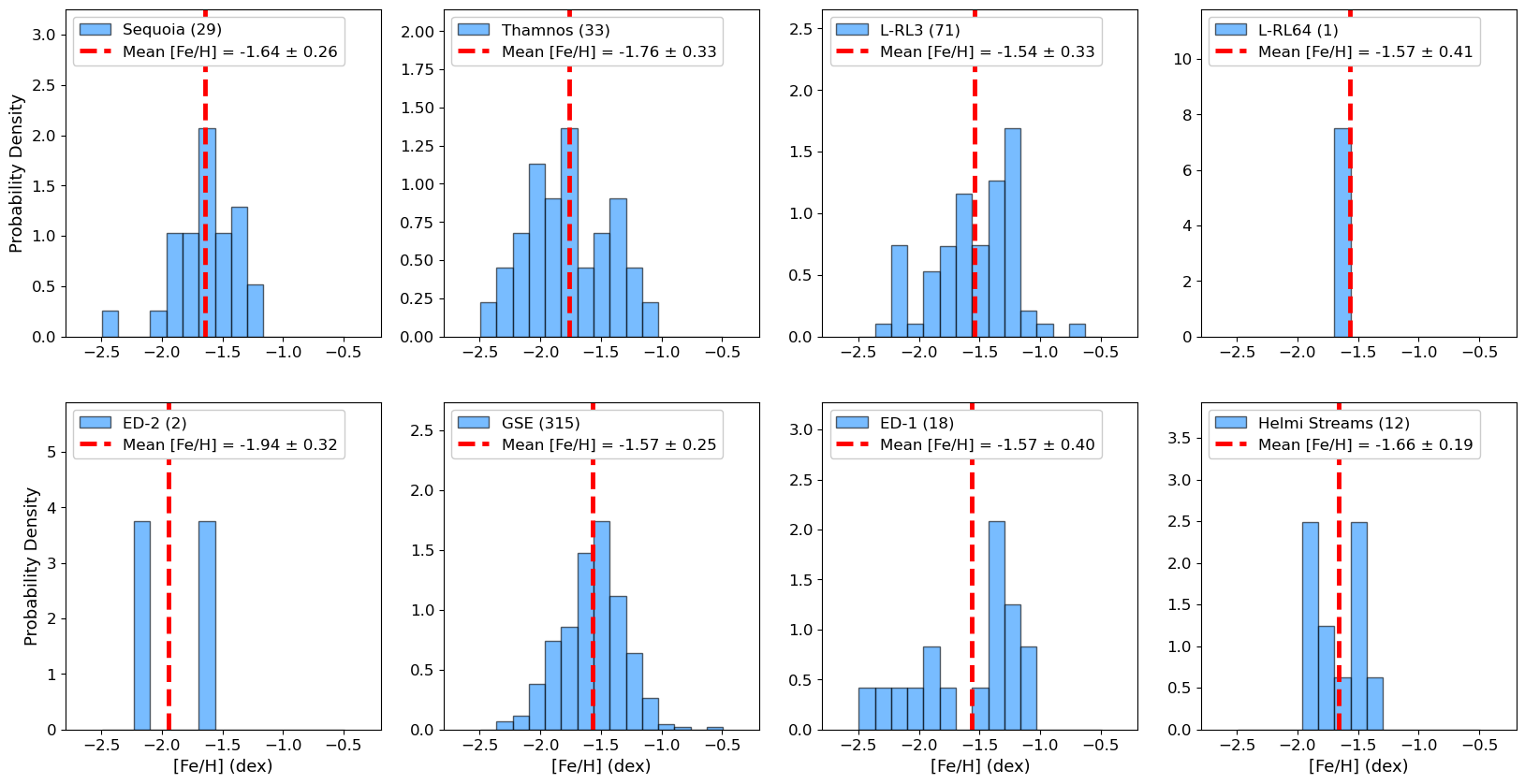}
\caption{Metallicity distribution of RRLs (light blue bins) in the known substructures of the MW halo. The dashed red line indicates the mean metallicity of RRLs in each substructure. Name of the substructure, number of RRLs with accurate metallicities, and mean metallicities are indicated in the legend.  The uncertainties are calculated as the weighted standard deviation of the mean. For substructures containing only one star, the uncertainty corresponds to the photometric metallicity uncertainty of that individual star.}\label{fig:histogram_known}
\end{figure*}

GSE is the remnant of relatively massive galaxy that merged with the MW approximately 8–11 Gyr ago \citep{Belokurov2018, Helmi2018}. This merger is considered a milestone in Galactic history and is responsible for a large fraction of the inner stellar halo \citep{DiMatteo2019, Gallart2019, Belokurov2020, Helmi2020}. 
One of the defining characteristics of the GSE population is the stars’ highly eccentric orbits and slightly retrograde motion \citep{Myeong2018, Koppelman2018, Haywood2018}. Various estimates of the stellar mass of the GSE progenitor exist in the literature, ranging from $2.7\times10^8$ to a few$\times10^9\ M_\odot$ \citep[e.g.][]{Helmi2018, vincenzo19, Kruijssen2020, Naidu2020, lane23}.


The CLiMB algorithm identified 398 RRLs in GSE, with accurate photometric metallicities from \citet{Muraveva2025} available for 315 of them. Fig.~\ref{fig:histogram_known} shows the metallicity distribution of RRLs in the GSE. The distribution appears approximately Gaussian, with a mean metallicity of [Fe/H] $= -1.57 \pm 0.25$~dex. The metallicity of old stellar populations, such as RRLs (age$>10$~Gyr), represents the chemical evolution state of the GSE progenitor at a very early epoch. On one side, the metallicity we derive is fully consistent with that reported by \citet{Ceccarelli2024}, who focused on the very retrograde, high-energy subset of GSE stars, which is expected to be dominated by stars originating from the outer regions of the GSE progenitor, where star formation likely ceased earlier than in the inner regions \citep{Koppelman2020,Skuladottir2025}. On the other side, the mean metallicity of RRLs in GSE structure is significantly lower than the mean photometric metallicities derived for GSE giant stars ([Fe/H]=$-1.19$~dex, \citealt{Bellazzini2023}) and also lower than the GSE metallicity derived from the Large Sky Area Multi-Object Fiber Spectroscopic Telescope (LAMOST; \citealt{Cui2012}) Low-Resolution Spectroscopic (LRS) Survey ([Fe/H]=$-1.24$~dex, D23). Such observed difference in mean metallicities between the RRLs found in this study and the overall stellar population of the GSE progenitor from the literature suggests that the GSE progenitor continued to form stars for an extended period after the formation epoch of the RRLs \citep[see e.g.][]{Gallart2019,Gonzalez-koda2025}, until star formation was eventually halted by the structure's accretion onto the MW. 

\subsection{Thamnos}\label{subsec:thamnos}

The Thamnos structure was discovered by \citet{Koppelman2019a} in the local retrograde halo using data from {\it Gaia} DR2 \citep{Brown2018}, the extended Apache Point Observatory Galactic Evolution Experiment (APOGEE) DR14 \citep{Abolfathi2018}, the LAMOST \citep{Cui2012}), and the Radial Velocity Experiment (RAVE) DR5 \citep{Kunder2017}. \citet{Koppelman2019a} identified two over-densities in chemo-dynamical space (Thamnos 1 and Thamnos 2), which they attributed to a common progenitor with a mass of less than $5 \times 10^6 \ M_\odot$. Thamnos is thought to be the remnant of a small dwarf galaxy accreted at an early stage, when the MW was not yet massive, or possibly around the same time as the GSE event \citep{Naidu2020}. \citet{Naidu2020} analysed the metallicity distribution of stars in Thamnos using data from the Hectochelle in the Halo at High Resolution (H3) survey \citep{Conroy2019}, and found a strong peak at [Fe/H] = $-1.9$~dex, which they attributed to Thamnos, as well as a second peak at [Fe/H] $\sim -1.2$~dex, attributed to GSE contamination. Using the MZR, \citet{Naidu2020} estimated a stellar mass of $2 \times 10^6 \ M_\odot$ for Thamnos, in good agreement with the earlier estimate by \citet{Koppelman2019a}. 

Similarly, \citet{Dodd2025b}, by analysing the Thamnos metallicity distribution from LAMOST DR7 \citep{Cui2012}, found that Thamnos has a pronounced metal-poor tail around [Fe/H]$\sim-2$~dex, which they attributed to the signature of a small dwarf galaxy. \citet{Dodd2025b} also reported a large peak in the metallicity distribution around [Fe/H]$\sim-1.4$~dex, which they associated with a combination of GSE contamination and the ancient, {\it in-situ} population of Aurora \citep{Belokurov2022}. \cite{ceccarelli25} estimated a contamination fraction for the \citet{Dodd2025b} sample of Thamnos of about 78\% across the whole metallicity range, mostly coming from {\it in-situ} stars. \citet{Mori2025} found the fractions of stars in Thamnos chemically compatible with GSE and the metal-poor disc to be 48\% and 19\%, respectively.

Using the CLiMB algorithm, we identified 48 RRLs in Thamnos. Fig.~\ref{fig:histogram_known} shows the metallicity distribution for the 33 RRLs in Thamnos with accurate metallicity estimates. The distribution appears visually bimodal, in agreement with previous studies \citep{Naidu2020, Dodd2025b}. We modelled this bimodal metallicity distribution with a two-component Gaussian Mixture Model (GMM). This approach provides a practical, data-driven method to statistically decompose the observed distribution into two Gaussian components, each representing a distinct subpopulation. The GMM fitting divides the Thamnos sample into a metal-poor (23 RRLs) and a metal-rich (10 RRLs) component. The corresponding mean metallicities are [Fe/H] = $-1.94 \pm 0.20$~dex and [Fe/H] = $-1.35 \pm 0.12$~dex, respectively \footnote{Mean metallicities derived directly from the two-component GMM fitted to the [Fe/H] distribution for metal-poor and metal-rich components are $-1.94\pm0.21$ and $-1.36\pm0.11$~dex, respectively.}. We associate the metal-poor component with the ``true'' Thamnos population, corresponding to the actual remnant of the accreted dwarf, and identified independently by \cite{Dodd2025b} using {\it Gaia} photometry and by \cite{ceccarelli25} using HR-spectroscopy. The metal-rich component is likely contamination from RRLs of GSE and \textit{in-situ} origin. Interestingly, the RRL number ratio between the two populations favours the ``true''  Thamnos, whereas in all other studies the likely contaminating, more metal-rich component is dominant. RRLs may therefore be ideal tracers for selecting the ``true'' Thamnos population.


\subsection{Sequoia}\label{subsec:sequoia}

Sequoia is a significant substructure in the local retrograde halo at higher energies \citep{Myeong2019, Matsuno2019}. 
Its true nature has been debated for some time, either as an independent merger event, or as the most external population of GSE lost during its first pericentric passage around the MW \citep{Koppelman2020}. Chemical evidence supports its interpretation as a separate dwarf galaxy \citep{matsuno22, Ceccarelli2024}, having stellar mass estimates ranging from $M_\star \sim 10^7 M_\odot$ \citep{Koppelman2019a, Helmi2000} to $M_\star = (0.8 \pm 0.2) \times 10^8 M_\odot$ \citep{Kruijssen2020}. 



We analysed 29 RRLs with accurate metallicities associated with Sequoia using the CliMB algorithm. Their Metallicity Distribution Function (MDF) is shown in Fig.~\ref{fig:histogram_known}. 
The mean metallicity of RRLs in Sequoia is [Fe/H] = $-1.64 \pm 0.26$ dex. This value is in good agreement with the mean spectroscopic metallicities derived for the overall Sequoia stellar population from the LAMOST LRS survey ([Fe/H]=$-1.56$ dex, \citealt{Dodd2023}) and HR spectroscopic data obtained with the Very Large Telescope and the Large Binocular Telescope ([Fe/H] = $-1.66$ dex, \citealt{Ceccarelli2024}). As discussed in Section~\ref{subsec:GSE}, a significant difference between the mean metallicities of RRLs and the overall stellar population in GSE could indicate that the GSE progenitor continued to form stars for an extended period after the epoch of RRL formation (>10~Gyr), before star formation was halted by the accretion of GSE by the MW. If we assume that the progenitors of GSE and Sequoia are two independent galaxies, the small difference between the mean metallicity of RRLs and the overall stellar population in Sequoia suggests that star formation in this galaxy stopped close to the epoch of RRL formation. This early cessation of star formation could be due to the accretion of Sequoia by the MW occurring before the MW accreted GSE. Alternatively, star formation could have stopped earlier due to the smaller mass of the Sequoia progenitor compared to GSE, leading to less efficient and shorter-lived star formation. If we assume instead that Sequoia and GSE originated from a single progenitor, their slight chemical difference ([Fe/H] of $-1.57$ dex for GSE versus $-1.64$ dex for Sequoia) could reflect an internal metallicity gradient of the progenitor galaxy. In this scenario, Sequoia represents the outer metal-poor envelope that was tidally stripped early during the merger.
In contrast, the GSE component represents the progenitor core, which remained gas rich and continued forming stars and thus enriching until final accretion.

\subsection{Helmi streams}\label{subsec:helmi}

The Helmi streams were among the first bona fide accreted substructures discovered in the MW halo via dynamical analysis \citep{Helmi1999}. Subsequently, \citet{Koppelman2019a, Koppelman2019b} provided an updated view of these streams using {\it Gaia} DR2 data. The Helmi streams are interpreted as the debris of a disrupted dwarf galaxy, with the time of accretion estimated to be 5–8 Gyr ago \citep{Kepley2007, Koppelman2019b}.
The stellar mass of the progenitor estimated in the literature is in the range $M_\star = (0.5$–$1) \times 10^8 M_\odot$ \citep[e.g.][]{Koppelman2019b, Naidu2020, Kruijssen2020}.

\citet{Bellazzini2023} found that the Helmi streams have a metallicity distribution characterized by a median of $-1.52$~dex, with the 16th and 84th percentiles at $-1.87$~dex and $-1.21$~dex, respectively. The MDF of the 12 RRLs in the Helmi streams identified by the CLiMB algorithm is shown in Fig.~\ref{fig:histogram_known}. Their metallicities span a range from -1.90 to -1.42~dex, with a mean of [Fe/H]$=-1.66 \pm 0.19$~dex, which is in good agreement with the value reported by \citet{Bellazzini2023}. As in the case of Sequoia (Section \ref{subsec:sequoia}), this could indicate that star formation in the Helmi streams progenitor stopped close to the epoch of RRL formation.

\subsection{ED-1/L-RL3}\label{subsec:ed1_lrl3}

\begin{figure*}
\includegraphics[width=18cm]{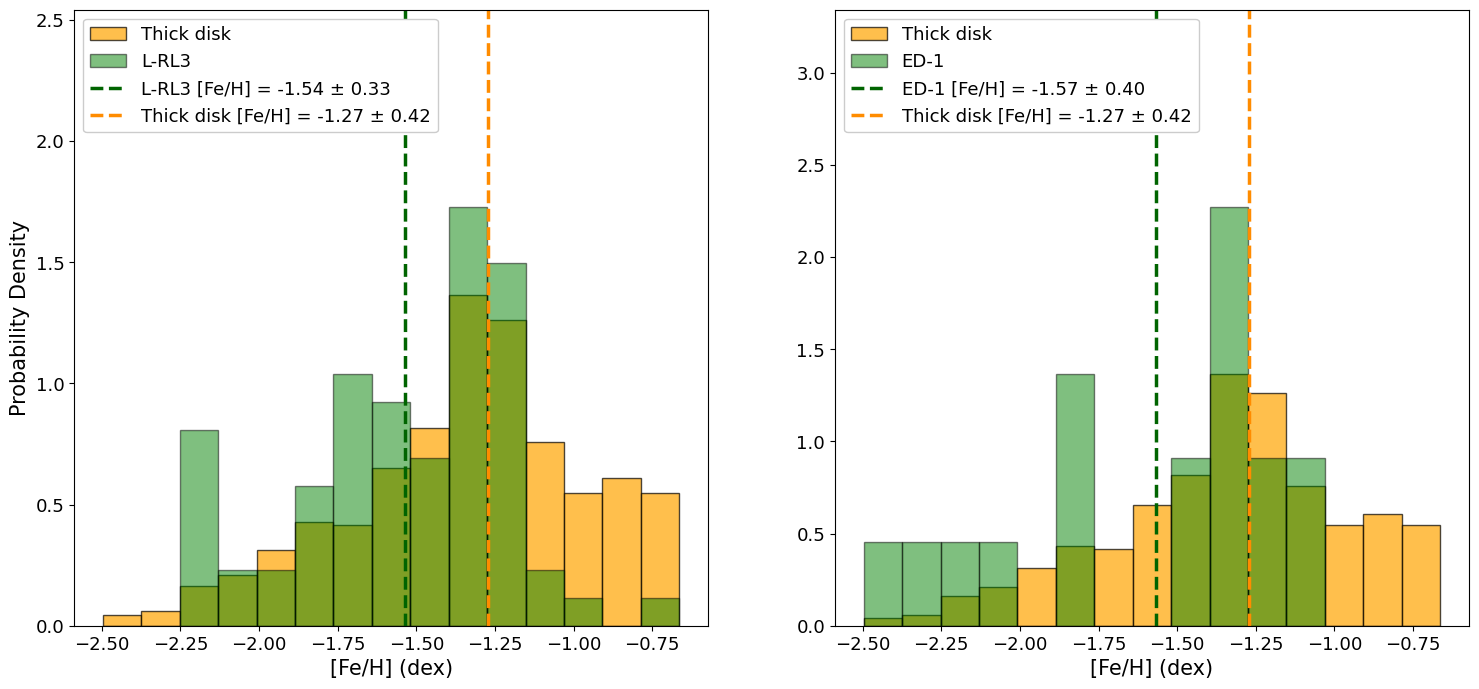}
\caption{Metallicity distribution of RRLs in L-RL3 ({\it left panel}) and ED-1 ({\it right panel}) shown in green bins. For comparison, the metallicity distribution of thick disc RRLs, defined in Section~\ref{subsec:disk}, is shown in orange bins. Dashed green and orange lines indicate the mean metallicities of the substructures and the thick disc, respectively.}\label{fig:ed1_lrl3}
\end{figure*}

ED-1 is a newly identified dynamical group discovered by D23. The MDF of ED-1, as studied by D23, spans a broad range and exhibits multiple peaks, approximately corresponding to the HD, GSE, and a more metal-poor component at [Fe/H] $\sim -1.8$~dex. D23 suggested that ED-1 includes contamination from both the HD and GSE, but the presence of a low-metallicity peak in its MDF, together with the chemical abundances of some of its stars, points toward an accreted origin for at least part of this substructure.

L-RL3 is a prominent substructure in the local stellar halo of the MW, located at low energy in IoM space. It was first identified by \citet{Lovdal2022} and \citet{Ruiz-Lara2022}, and later further characterized by D23. Its MDF reveals two main components: a high-metallicity population similar to that of the HD, and a well-populated low-metallicity tail. D23 suggested that L-RL3 comprises both {\it in-situ} and accreted stars.

Accurate metallicities are available for 71 RRLs in L-RL3 and 18 RRLs in ED-1, with mean values of [Fe/H] $= -1.54 \pm 0.33$~dex and [Fe/H] $= -1.57 \pm 0.40$~dex, respectively. 
Both substructures are relatively metal-rich and located close to the thick disc in IoM space (Fig.~\ref{fig:clusters_all_old}), which could indicate that they are heavily contaminated by stars of the thick disc/HD.
In Fig.~\ref{fig:ed1_lrl3}, we show the metallicity distributions of RRLs in L-RL3 (left panel) and ED-1 (right panel). For comparison, we also plot the metallicity distribution of RRLs in the thick disc, as defined by their eccentricity and $Z_\mathrm{max}$  (Section~\ref{subsec:disk}). The distributions appear visually similar, with closely positioned peak metallicities. The MDFs of both substructures also exhibit weak metal-poor tails, which may be remnants of small accreted systems.  We performed two-sample Kolmogorov–Smirnov (KS) tests to evaluate the hypothesis that L-RL3 and ED-1 are drawn from the same distribution as the thick disc RRLs. The resulting $p$-values are $7.2 \times 10^{-6}$ and $3.0 \times 10^{-2}$ for L-RL3 and ED-1, respectively. These results indicate that L-RL3 and ED-1 are statistically inconsistent with being drawn from the thick disc RRL distribution.

It is worth noting that, as mentioned in Section~\ref{subsec:comp}, there is a difference of more than 0.1~dex between the  mean metallicities of RRLs in ED-1 identified by cross-match with D23 and those identified using the CLiMB algorithm. Moreover, the CLiMB algorithm added 286\% more RRLs to ED-1 (see Table~\ref{tab:clusters_dodd}), compared with RRLs assigned to this substructure by D23. For a sanity check, we performed a two-sample KS test between the thick disc RRLs and RRLs classified as belonging to ED-1 only by D23. The resulting $p$-values is $4.2 \times 10^{-1}$, indicating no statistically significant difference between the two distributions. This suggests that the ED-1 sample from D23 is consistent with being drawn from the thick disc population. We caution that robust conclusions on the origin of ED-1 and L-RL3 cannot be drawn due to small sample sizes and the sensitivity of KS tests to the sample selection method. Still, the overlap in IoM space and metallicity distributions indicates that both substructures likely contain significant contamination from thick disc stars, while the weak metal-poor tails in their MDFs may be signatures of remnants from small accreted systems.

\subsection{ED-2}\label{subsec:ed2}

Recently, D23, using {\it Gaia} DR3 data, discovered a new structure ED-2  in the retrograde halo within the region associated with Sequoia in IoM space. \citet{Dodd2025a} found that stars in ED-2 are very metal-poor (mean [Fe/H] $= -2.43$~dex) and exhibit a small intrinsic metallicity dispersion ($\sigma = 0.13$~dex), in agreement with previous studies \citep{Balbinot2024, Ceccarelli2024}. They also estimated the age of ED-2 to be comparable to the age of the Universe, consistent with \citet{Balbinot2024}. The progenitor of ED-2 is likely an ancient star cluster \citep{Balbinot2024, Dodd2025a}.

The CLiMB algorithm identified three RRLs in ED-2. Accurate metallicity estimates were available for only two of them, with values of [Fe/H] $= -2.21\pm0.34$~dex and [Fe/H] $= -1.57\pm0.39$~dex, resulting in a mean metallicity of [Fe/H] $= -1.94 \pm 0.32$~dex for ED-2. Fig.~\ref{fig:histogram_known} shows the metallicity distribution of RRLs in ED-2.
The observed spread in metallicities, which differs from values reported in the literature, could be due to contamination of the ED-2 RRL sample by a more metal-rich star of {\it in-situ} or GSE origin. Indeed, as shown in Table~\ref{tab:clusters_dodd}, only one star with [Fe/H] $= -2.21\pm0.34$~dex was classified as belonging to ED-2 based on a cross-match with the D23 catalogue.
We thus conclude that the additional metal-rich star identified in ED-2 by the CLiMB algorithm may not be a true member of the substructure and is likely a contaminant. In this case, the metallicity of the single confirmed ED-2 star would be consistent with literature values, supporting the interpretation of ED-2 as an extremely metal-poor structure, possibly the remnant of a GC. 

\subsection{L-RL64/Antaeus}\label{sec:lrl64}

L-RL64 is a newly discovered cluster located in the most retrograde region of IoM space \citep{Lovdal2022, Ruiz-Lara2022}. Subsequently, \citet{Oria2022} independently identified this substructure, naming it Antaeus. L-RL64 was initially thought to be the extreme retrograde tail of the Sequoia merger event, but it is now considered an independent accretion event, unrelated to Sequoia \citep{Oria2022, Ruiz-Lara2022, Dodd2023, Ceccarelli2024}. 

Only one RRL from our sample with [Fe/H] $= -1.57 \pm 0.41$~dex was identified as belonging to L-RL64, both by the CLiMB algorithm and via cross-matching with the D23 catalogue. Given its large uncertainty, the metallicity of this star is consistent with the distribution observed for L-RL64, as found by \citet{Ceccarelli2024}, which peaks at [Fe/H] $= -1.83$ dex with a dispersion of 0.27 dex. However, we cannot exclude the possibility that the only RRL star identified in L-RL64 is a contaminant of {\it in-situ} or GSE origin.


\subsection{Shiva and Shakti}\label{subsec:shiva_shakti}

\begin{figure}
\includegraphics[width=\columnwidth]{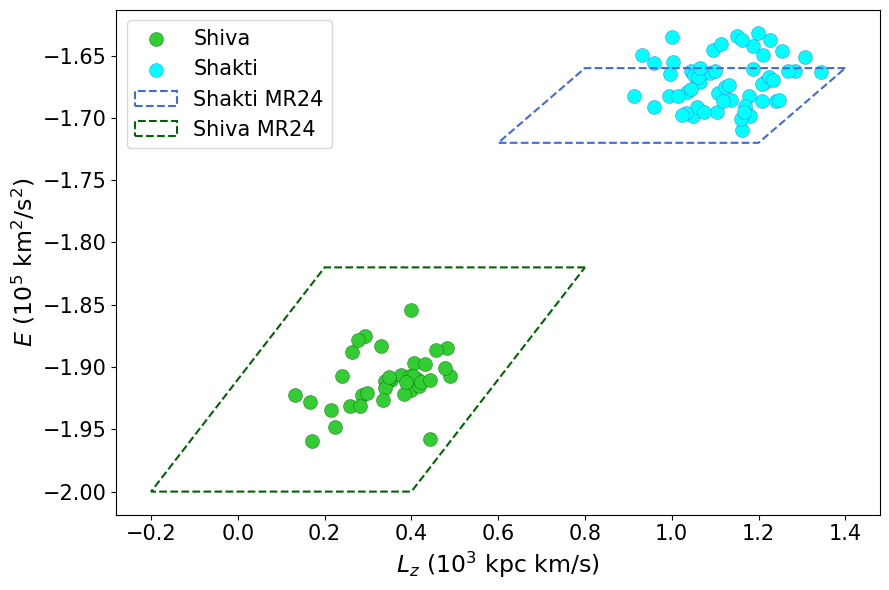}
\caption{RRLs in Shiva (green dots) and Shakti (cyan dots) identified by the CLiMB algorithm. Dashed green and blue lines outline the regions used by \citet[MR24]{Malhan2024} to identify members of Shiva and Shakti, respectively.}\label{fig:shiva_shakti_area}
\end{figure}

\begin{figure*}
\includegraphics[width=18cm]{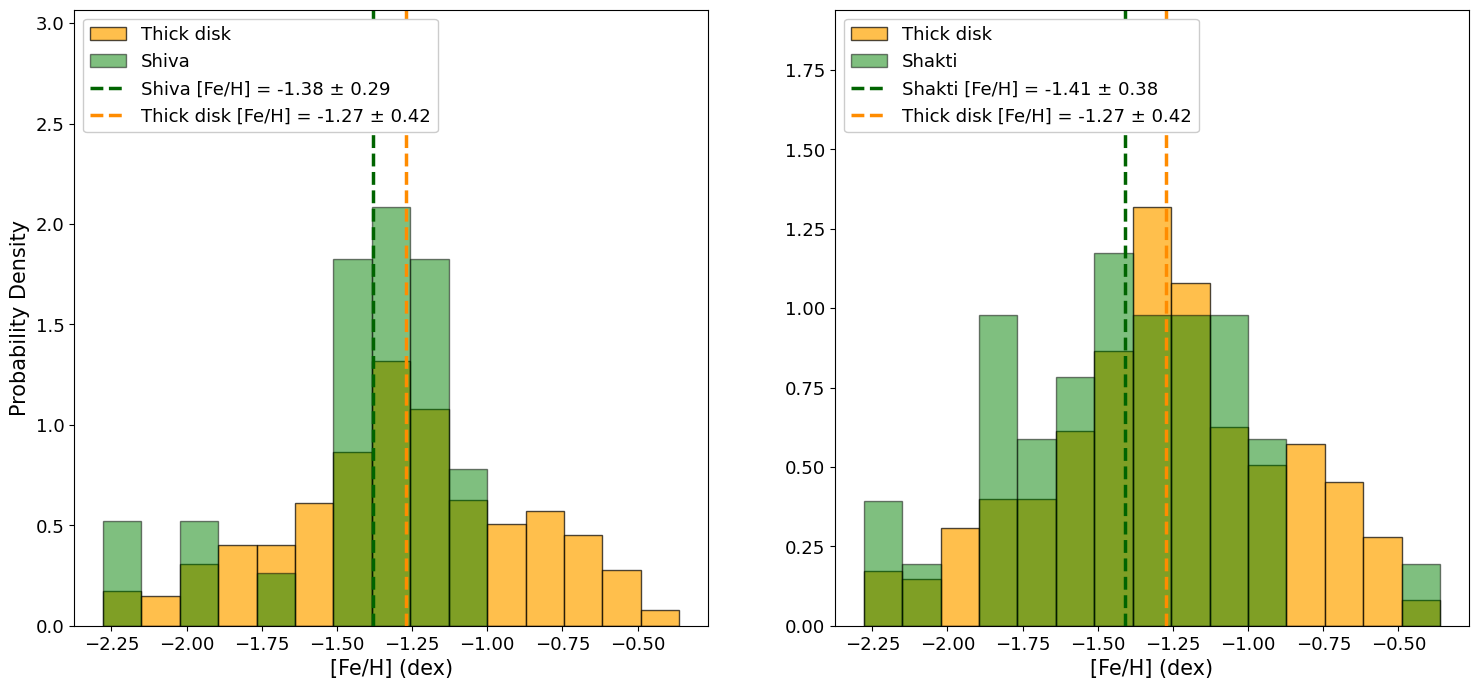}
\caption{Metallicity distribution of RRLs in Shiva ({\it left panel}) and Shakti ({\it right panel}) shown in green bins. For comparison, the metallicity distribution of thick disc RRLs, defined in Section~\ref{subsec:disk}, is shown in orange bins. Dashed green and orange lines indicate the mean metallicities of the substructures and the thick disc, respectively.}\label{fig:shakti_met}
\end{figure*}

\begin{figure}
\includegraphics[width=\columnwidth]{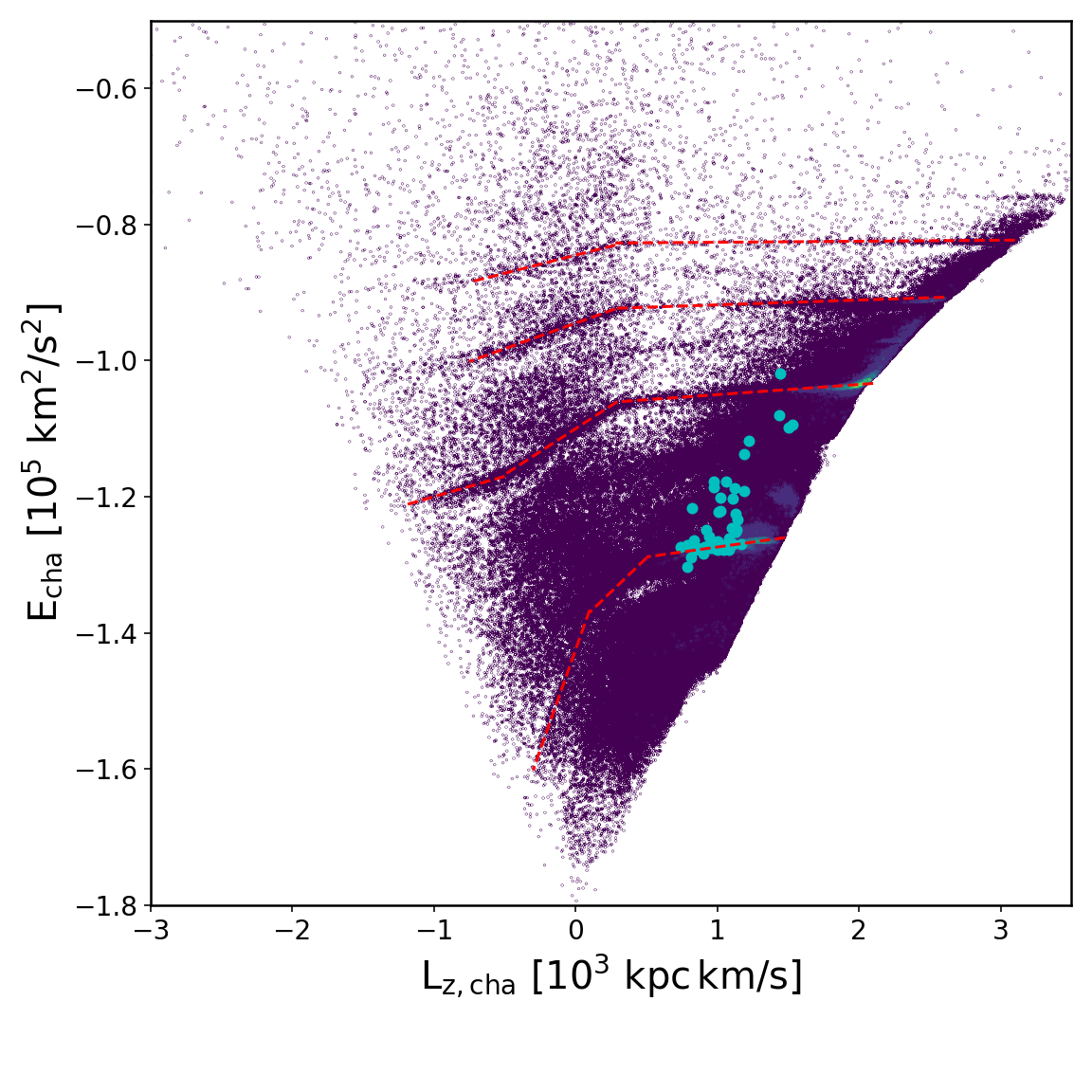}
\caption{Distribution of stars from \citet{Bellazzini2023} used to identify the resonant loci on the characteristic energy versus angular momentum plane (purple dots). Cyan points represent RRLs identified in Shakti by CLiMB, while red dashed lines indicate the resonance loci.
 }\label{fig:shakti_res}
\end{figure}


\begin{figure*}
\includegraphics[width=17cm]{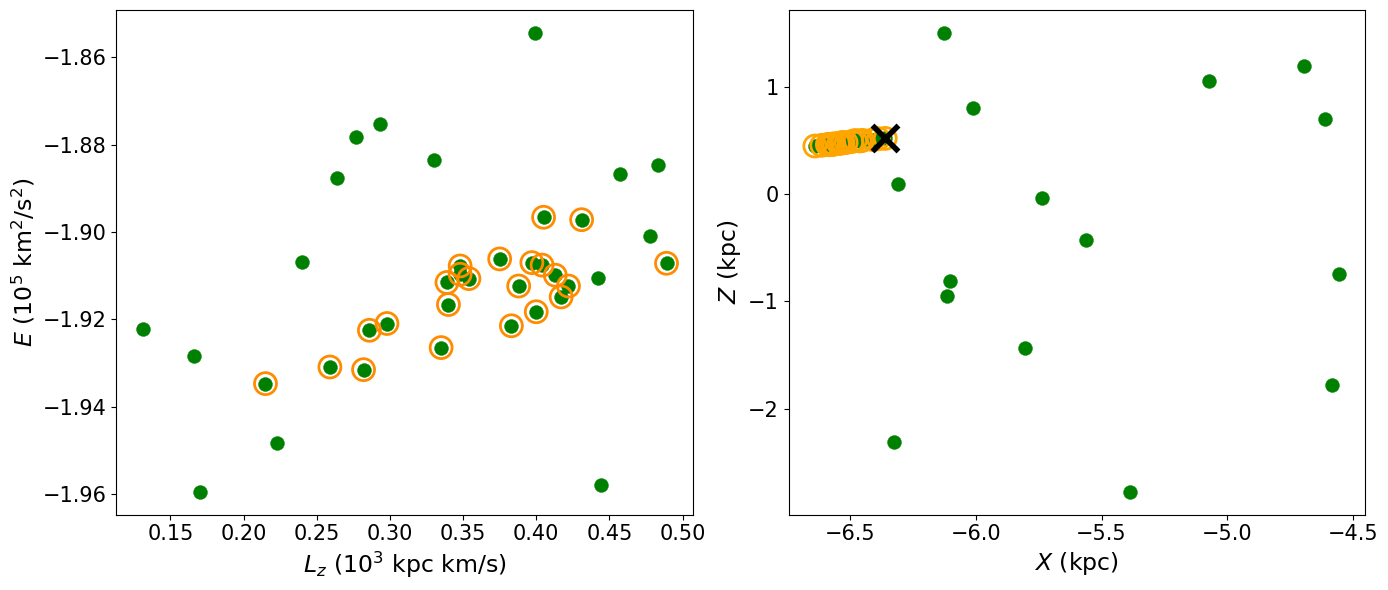}
\caption{Distribution of RRLs in Shiva (green dots) shown in the energy versus $L_z$ plane ({\it left panel}) and the Cartesian 
$Z$ versus $X$ plane ({\it right panel}). Orange open circles indicate RRLs identified as members of NGC~6121, according to the \citet{Clement2001} catalogue. The black cross identifies the centre of NGC~6121, calculated using coordinates from \citet{Baumgardt2019} and the distance from \citet{Baumgardt2021}.}\label{fig:ngc6121}
\end{figure*}

With the explorative phase of the CLiMB algorithm, we identified two substructures not included in the D23 catalogue, located near the MW disc (yellow and pink dots in Fig.~\ref{fig:clusters_all_new}). Recently, \citet{Malhan2024} reported the discovery of two new substructures in the MW halo, named Shiva and Shakti. In Fig.~\ref{fig:shiva_shakti_area}, we show the positions of RRLs in our newly identified substructures on the $E$–$L_z$ plane in correspondence to the area used to outline members of Shiva and Shakti by \citet{Malhan2024}, who also adopted the MW potential of \citet{McMillan2017}. As can be seen, there is significant overlap. We therefore conclude that the two substructures identified by the CLiMB algorithm correspond to Shiva and Shakti discovered by \citet{Malhan2024}, although we cannot be entirely certain, given that our sample of 4933 RRLs is significantly smaller than the sample of stars from {\it Gaia} DR3 used by \citet{Malhan2024} to identify these over-densities (5,799,724 stars).

\citet{Malhan2024} analysed various hypotheses regarding the origin of these two substructures. They found that both exhibit orbit-space distributions indicative of an accreted origin, yet their chemical abundance patterns are more typical of an {\it in-situ} population. \citet{Malhan2024} proposed two possible scenarios. In the first, these prograde substructures may have formed through resonant orbit trapping of field stars by the rotating Galactic bar. This interpretation is consistent with the findings of \citet{Dillamore2023}, who identified a prominent ridge feature in phase space at constant energy and positive $L_z$, likely originating from stars trapped in corotation resonances with the MW’s bar. \citet{Dillamore2023} showed that this ridge contains both metal-rich and metal-poor stars, with orbits lying between typical disc-like and halo-like trajectories. \citet{Malhan2024} noted that Shakti is located near the tip of this ridge in the $E-L_z$ plane, suggesting that it could have a resonant origin. In the second scenario, proposed by \citet{Malhan2024}, Shiva and Shakti are interpreted as protogalactic fragments that underwent rapid star formation and merged early in the Galaxy’s history, similar to the building blocks of the MW’s ancient inner regions.

The CLiMB algorithm identified 58 RRLs as belonging to Shakti, of which 40 have accurate metallicity measurements. Their metallicity distribution is shown in the right panel of Fig.~\ref{fig:shakti_met}, compared to the metallicity distribution of the thick disk, as defined in Section~\ref{subsec:disk}. As shown, the two distributions are visually similar.
According to a KS test, the probability that our RRL samples of Shakti and the thick disc are drawn from the same metallicity distribution is not negligible ($p = 0.16$).



To investigate the hypothesis of a resonant origin for the over-density identified as Shakti, we implemented the approach of \citet{DeLeo2025} to check whether any Shakti members were trapped in resonant orbits. Briefly, taking the presence and influence of the rotating bar of the MW fully into account, \citet{DeLeo2025} developed an empirical method to identify resonant loci in dynamical parameter spaces. Analysing the large sample of stars from \citet{Bellazzini2023} composed of giants from the {\it Gaia} Synthetic Photometry Catalogue \citep{Montefriffo2023}, the loci appear as clear over-densities in the periocentric radius–eccentricity space and in the space of "characteristic" energy ($E_{cha}$) and angular momentum ($L_{z,cha}$). These characteristic values are defined as the mean of the corresponding IoM across the entire orbit integration \citep{Moreno15, Moreno21}. Stars located close to or on top of the resonant loci in both dynamical spaces are considered trapped by resonances induced by the rotating bar. Fig.~\ref{fig:shakti_res} shows the RRLs of Shakti (cyan points) in $L_{z,cha}$--$E_{cha}$ space, over-plotted on the distribution of stars from \citet{Bellazzini2023} used to identify the resonant loci, which are highlighted with the red lines. As can be seen, Shakti lies precisely along the resonant loci in the $L_{z,cha}$--$E_{cha}$ plane.  We found that 14 out of 58 RRLs in Shakti (24\%) are potentially trapped in resonant orbits, as evidenced by the cyan points clustered around $E_{cha} \approx -1.3 \times 10^5 \text{ km}^2\text{/s}^2$ and $L_{z,cha} \approx 10^3 \text{ kpc km/s}$.
Based on Shakti’s location in the $L_{z,cha}$--$E_{cha}$ plane, the metallicity distribution of its RRLs, which is similar to that of the thick disk, and the significant fraction of stars held in resonant orbits, we conclude that the substructure corresponding to Shakti is likely part of the thick disk population that became trapped in a ridge-like feature due to resonances with the MW bar. 

Regarding the Shiva substructure, we identified 38 RRLs within it using the CLiMB algorithm. Photometric metallicity estimates are available for 30 of them, yielding a mean metallicity of [Fe/H] = $-1.38 \pm 0.29$~dex. Their MDF is shown in the left panel of Fig.~\ref{fig:shakti_met}. Curiously, we found that 23 of the 38 RRLs discovered in correspondence with Shiva by CLiMB are members of the GC NGC~6121 (M4). To identify RRLs associated with NGC~6121, we used the catalogue by \citet{Clement2001}\footnote{\url{https://www.astro.utoronto.ca/~cclement/cat/listngc.html}}
, which is continuously updated based on the literature and summarizes the numbers and types of variable stars in various GCs. Fig.~\ref{fig:ngc6121} shows the RRLs identified by CLiMB as part of Shiva (green dots), with the RRLs associated with NGC~6121 over-plotted (orange open circles). The black cross indicates the Cartesian coordinates of the centre of the NGC~6121 cluster, calculated as described in Section~\ref{sec:data}, using the coordinates from \citet{Baumgardt2019} and the distance to NGC~6121 as determined by \citet{Baumgardt2021}. The discovery of Shiva by \cite{Malhan2024} came after excluding  the stars lying within five tidal radii of the GCs from the analysed sample. The Shiva over-density should therefore not coincide with NGC~6121. However, we cannot exclude that the stellar stream of NGC~6121, and especially the stars that have been lost recently by the GCs, might have contributed in enhancing the signal of such a dynamical over-density. It would be interesting to check this possibility by looking for the chemical signatures of GCs in the light element abundances of Shiva members. 

\section{Mass-metallicity relation}\label{sec:mzr}

\begin{figure*}
\centering
\includegraphics[width=15cm]{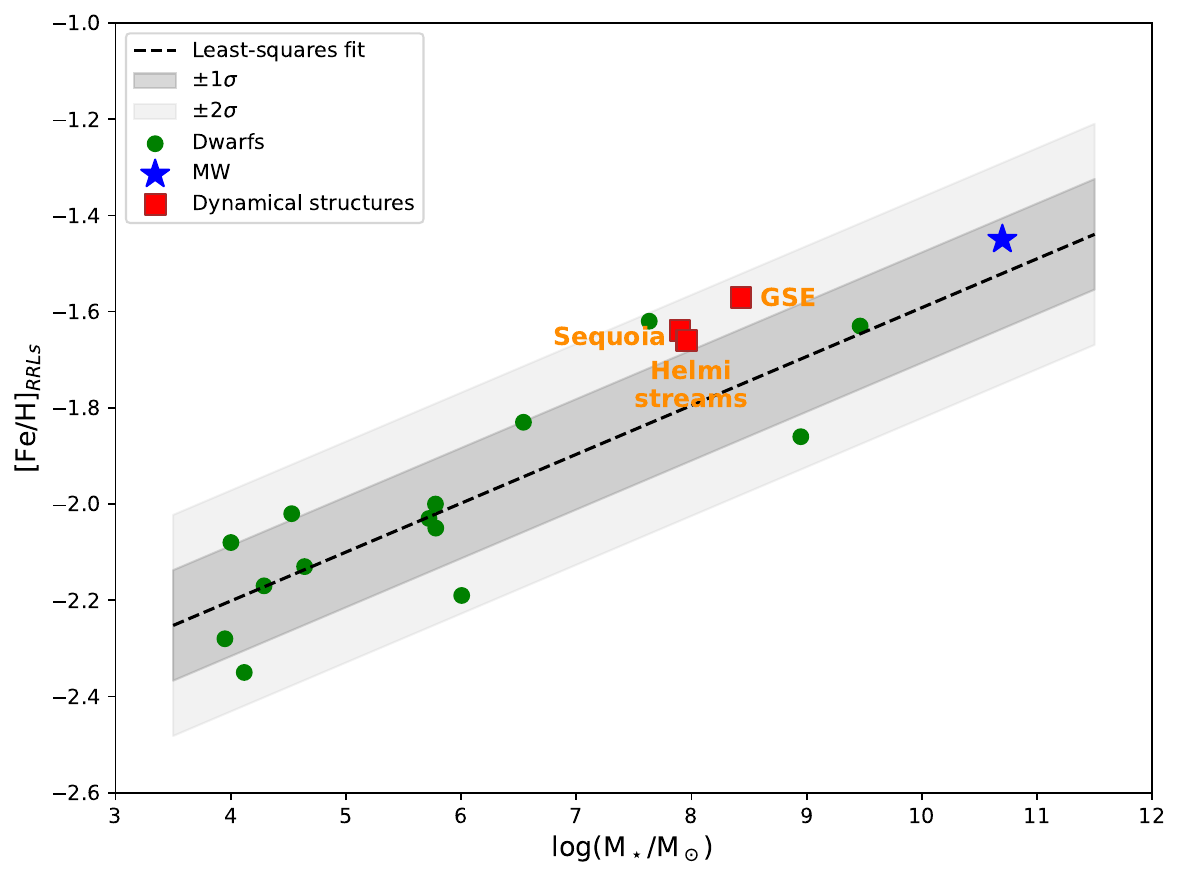}
\caption{Mean photometric metallicity of RRLs versus stellar mass for 14 dwarf galaxies (green circles) and the MW (blue star) from \citet{Bellazzini2025}, and for the known dynamical substructures analysed in this work (red squares). The dashed black line shows the best fit to the 14 dwarfs from \citet{Bellazzini2025}, while the dark and light grey areas indicate the $\pm 1\sigma$ and $\pm 2\sigma$ regions, respectively.}\label{fig:mzr}
\end{figure*}

The MZR, the correlation between a galaxy’s stellar mass and its metallicity, is a fundamental law of galaxies. The MZR has been observed to evolve with cosmic time. In \citet{Bellazzini2025}, we demonstrated that the metallicity distributions of RRLs can be used to investigate the MZR of galaxies at early epochs ($z \gtrsim 3$), close to when the progenitors of present-day dynamical substructures merged into the MW. To construct the MZR in \citet{Bellazzini2025}, we used the mean photometric metallicities of RRLs from \citet{Muraveva2025} in 14 dwarf galaxies and the MW, while stellar masses were taken from \citet{Pace2024}. 

In Fig.~\ref{fig:mzr}, we plot the mean metallicity versus stellar mass for 14 dwarf galaxies (green circles) and the MW (blue star) from \citet{Bellazzini2025}, as well as the known dynamical substructures analysed in this study (red squares): GSE, Sequoia, and the Helmi streams, for which stellar masses were determined by \citet{Kruijssen2020}. The stellar masses were estimated by means of the neural network using the ages, metallicities, and orbital properties of GCs associated with the same progenitor galaxy. As can be seen, all analysed dynamical substructures lie within $2\sigma$ of the best-fit relation based on the 14 dwarf satellites of the MW \citep{Bellazzini2025}, demonstrating that the MZR based on RRLs is a powerful tool for estimating the mass of both existing and accreted satellites of the MW.

The substructures identified by CLiMB in correspondence with Shiva and Shakti have the mean metallicities of [Fe/H] = $-1.38 \pm 0.29$~dex and [Fe/H] = $-1.41 \pm 0.38$~dex, respectively. The MZR based on RRLs would yield corresponding progenitor stellar masses of $M_\star = 5 \times 10^{11} M_\odot$ for Shiva and $M_\star = 3 \times 10^{11} M_\odot$ for Shakti, higher than those of the MW, supporting the conclusion that the groups of RRLs identified by CLiMB that we associate to Shiva and Shakti are likely dominated by stars originated {\it in-situ},  as already discussed in Section~\ref{subsec:shiva_shakti}.

Based on the MZR of RRLs and assuming ED-2 is a galaxy,  we estimate its approximate mass to be $M_\star = 10^4,M_\odot$ if we consider only the metal-poor star ([Fe/H] $= -2.21 \pm 0.34$ dex), which is in agreement with the findings of \citet{Balbinot2024}. If instead we consider the mean metallicity of the two RRLs in ED-2, [Fe/H] $= -1.94 \pm 0.32$ dex, the stellar mass would be $M_\star = 3 \times 10^6,M_\odot$, close to the Thamnos stellar mass \citep{Naidu2020}. However, as discussed in Section~\ref{subsec:ed2}, the more metal-rich RRL in ED-2 is likely a contaminant.
Moreover, as discussed in Section~\ref{subsec:ed2}, the progenitor of ED-2 is likely the remnant of a GC \citep{Balbinot2024, Dodd2025a}.

\section{Conclusion}\label{sec:concl}

We analysed a sample of 4933 RRLs for which accurate coordinates, radial velocities, and proper motions were available from the {\it Gaia} EDR3 \citep{Brown2021} and DR3 \citep{Vallenari2023} catalogues, while geometric distances were estimated from the {\it Gaia} EDR3 parallaxes using a Bayesian approach \citep{BJ2021}. We calculated the IoM ($E$, $L_z$, $L_\perp$) as well as orbital parameters (e.g. apocentre, pericentre, and eccentricity) for all 4933 RRLs in the sample. We then applied the semi-supervised, two-phase CLiMB algorithm (Monti et al., in prep.) to identify both previously reported by D23 and newly discovered dynamical substructures. In the first phase, CLiMB assigned 630 RRLs to known substructures identified by D23, representing a 35\% increase in membership compared to their original sample. The second, exploratory phase successfully identified the extended MW disk structure, as well as two over-densities in correspondence with the newly discovered substructures Shiva and Shakti \citep{Malhan2024}. Photometric metallicities with uncertainties smaller than 0.5~dex were available for 3614 of the 4933 RRLs. We used these metallicities to perform a detailed analysis of identified dynamical substructures.

\begin{itemize}
    \item {\it MW disk:} The CLiMB algorithm identified a disk-like structure that contains RRLs originating from the thin disk, thick disk, and HD. Based on their eccentricity and $Z_{\rm max}$, we separate the RRLs in this structure into a thin disc component, consisting of 238 stars, and a thick disc component, consisting of 600 RRLs. The two components have mean metallicities of [Fe/H] = $-0.75 \pm 0.40$~dex and [Fe/H] = $-1.27 \pm 0.42$~dex, respectively.

    \item {\it GSE:} The CLiMB algorithm identified 398 RRLs in GSE, 315 of which have accurate metallicity estimates from \citet{Muraveva2025}, with a mean metallicity of $[\text{Fe/H}] = -1.57 \pm 0.25$~dex.
    This value is significantly lower than the mean metallicity of GSE overall stellar population \citep{Dodd2023, Bellazzini2023}. This discrepancy suggests that star formation within the GSE progenitor continued for an extended period following the formation epoch of RRLs, persisting until the GSE structure was accreted by the MW.
    
    \item {\it Thamnos:} The RRL metallicity distribution in Thamnos was found to be bimodal. The metal-poor peak, likely representing the ``true'' Thamnos population, yielded a mean metallicity of $[\text{Fe/H}] = -1.94 \pm 0.20$ dex. The metal-rich component ($[\text{Fe/H}] = -1.35 \pm 0.12$ dex) is likely contamination from GSE and {\it in-situ} RRLs.
    
  \item {\it Sequoia and Helmi streams:} Sequoia RRLs exhibit a mean metallicity of $[\text{Fe/H}]= -1.64 \pm 0.26$ dex, while the mean metallicity of RRLs in the Helmi streams is $[\text{Fe/H}]= -1.66 \pm 0.19$ dex. Both substructures have mean RRL metallicities that are in good agreement with the values found for the overall stellar population in these substructures reported in the literature. This result suggests that, differently from GSE, star formation in these systems halted relatively close to the epoch of RRL formation.
    

    \item The mean metallicities of RRLs in ED-1 and L-RL3 are $[\mathrm{Fe/H}] = -1.57 \pm 0.40$~dex and $[\mathrm{Fe/H}] = -1.54 \pm 0.33$~dex, respectively. Based on their MDFs and their distribution in IoM space, we conclude that both substructures likely contain significant contamination from thick disc stars, while the weak metal-poor tails in their MDFs may be signatures of remnants from small accreted systems.

    \item {\it L-RL64 and ED-2:} Due to the small number of RRLs identified in L-RL64 (one star) and ED-2 (two stars with available metallicities), it is difficult to place reliable constraints on their origins. However, the single confirmed extremely metal-poor star in ED-2 ([Fe/H] $= -2.21 \pm 0.34$ dex) supports the hypothesis that it is a remnant of a GC. Meanwhile, we lack sufficient information to determine whether the single RRL discovered in L-RL64, with $[\mathrm{Fe/H}] = -1.57 \pm 0.41$ dex, is a true member or a contaminant of {\it in-situ} or GSE origin.

   \item {\it Shiva:} The exploratory phase of CLiMB led to the identification of two groups corresponding to the recently reported substructures Shiva and Shakti \citep{Malhan2024}. Our analysis suggests that the over-density of RRLs identified by the CLiMB algorithm in the region of Shiva is primarily due to RRLs belonging to the GC NGC~6121 (M4).
   
    \item {\it Shakti:} We found that Shakti lies precisely within a known ridge-like structure in the $L_{z,cha}$--$E_{cha}$ plane that originates from resonances with the MW bar. Moreover, we find that 14 out of 58 RRLs in Shakti (24\%) are trapped in resonant orbits. Finally, a KS test between the metallicity distributions of Shakti and the thick disc yields a $p$-value of 0.16, suggesting that the probability that our RRL samples of Shakti and of the thick disc are drawn from the same metallicity distribution is not negligible. We thus conclude that Shakti is likely part of the thick disk population that became trapped in a resonant feature.
    
\end{itemize}

Finally, we used the MZR based on RRLs \citep{Bellazzini2025} to test the nature of the identified dynamical substructures. RRLs are a valuable tool for analysing the MZR at early epochs ($z \gtrsim 3$), close to when the progenitors of present-day dynamical substructures merged into the MW. The known accreted substructures (GSE, Sequoia, Helmi streams) were found to lie within $2\sigma$ of the best-fit MZR derived from dwarf satellites. 

The high metallicities of clusters identified in correspondence with Shiva and Shakti conflict with an accreted origin according to the MZR, confirming their likely {\it in-situ} origin. Applying the MZR to the confirmed metal-poor RRL in ED-2, we find a progenitor mass of $M_\star \approx 10^4,M_\odot$, consistent with ED-2 having been a very low-mass galaxy. However, the progenitor of ED-2 is likely the remnant of a GC \citep{Balbinot2024, Dodd2025a}. Overall, this work demonstrates the power of using RRLs to refine our understanding of the origins and compositions of MW halo substructures. 

A more extensive study of MW dynamical substructures, as traced with RRLs, will be enabled by the arrival of {\it Gaia} Data Release 4 (DR4), currently expected in December 2026. Based on 66 months of observations, {\it Gaia} DR4 will provide high-cadence light curves and updated Fourier parameters for a vast number of RRLs, which will yield more accurate photometric metallicities, crucial for confirming the nature of ancient structures. Moreover, the accurate astrometric data will be further enhanced by a dramatic increase in radial velocities, delivering the 6D phase-space information required for sophisticated chemo-dynamical analysis of the MW halo. This comprehensive dataset will provide a timely opportunity to refine the CLiMB algorithm by optimizing its training sets to reduce contamination between overlapping structures. Together, these observational and algorithmic advancements stand to provide a significant contribution toward reconstructing the MW’s assembly history and characterizing its earliest building blocks.

\begin{acknowledgements}
We thank the anonymous referee for their thoughtful and constructive review which helped improve the paper. We also thank Dr. Michele Bellazzini and Dr. Emma Dodd for fruitful and interesting discussions which helped improve the manuscript.
This work made use of Astropy:\footnote{http://www.astropy.org} a community-developed core Python package and an ecosystem of tools and resources for astronomy \citep{astropy:2013, astropy:2018, astropy:2022}.
Support to this study has been provided by INAF Mini-Grant (PI: Tatiana Muraveva), by the Agenzia Spaziale Italiana (ASI) through contract ASI 2018-24-HH.0, its Addendum 2018-24-HH.1-2022 and contract ASI 2025-10-H.00, and by Premiale 2015, MIning The Cosmos - Big Data and Innovative. Italian MDL acknowledges financial support from the project LEGO – Reconstructing the building blocks of the Galaxy by chemical tagging (PI: Mucciarelli) granted by the Italian MUR through contract PRIN2022LLP8TK\_001. This work uses data from the European Space Agency mission {\it Gaia} (https://www.cosmos.esa.int/gaia), processed by the {\it Gaia} Data Processing and Analysis Consortium (DPAC; https: //www.cosmos.esa.int/web/gaia/dpac/consortium). Funding for the DPAC has been provided by national institutions, in particular the institutions participating in the {\it Gaia} Multilateral Agreement.

\end{acknowledgements}

\bibliographystyle{aa}
\bibliography{bibl}

\begin{appendix}
\onecolumn
\section{Dataset}\label{ap:dataset}

\begin{table*}[ht!]
\small
\centering
\caption{Parameters of the 4933 RRLs in the reference sample.}
\label{tab:gen} 
\begin{tabular}{cccccccccccc} 
\hline 
source\_id & RA & Dec.& \texttt{pmra} & \texttt{pmdec} & $L_z$ & $E$  & $L_\perp$ & $ecc$ & [Fe/H] &  $\sigma_{\rm [Fe/H]}$ & Substructure\\
\hline
5930541561668628736 & 249.1970 & -54.2037 & -17.171 & -14.857 & -745 & -177885 & 344 & 0.59 & -2.11 & 0.42 & Thamnos \\
6471410229148825088 & 310.1242 & -53.2819 & -7.662 & -28.896 & -1789 & -133202 & 821 & 0.56 & -1.75 & 0.42 & Sequoia \\
34113623857898880 & 44.5915 & 15.6717 & 1.760 & -37.565 & 98 & -135829 & 26 & 0.98 & -1.64 & 0.44 & GSE \\
1193183528770580224 & 238.6171 & 15.3561 & -17.853 & 1.871 & -252 & -154345 & 210 & 0.91 & -1.49 & 0.44 & GSE \\
1196737700107586176 & 238.3152 & 17.5699 & -11.729 & -4.870 & -533 & -179584 & 361 & 0.59 & -1.88 & 0.41 & Thamnos \\
2098060604631085952 & 279.6238 & 39.5011 & -1.520 & -5.618 & 291 & -180257 & 503 & 0.84 & -1.65 & 0.35 & L-RL3 \\
5768557209320424320 & 248.1040 & -83.9035 & -58.488 & -120.550 & -218 & -160786 & 789 & 0.91 & -1.97 & 0.50 & GSE \\
702294777158642688 & 139.9645 & 33.8732 & -3.352 & -12.630 & 226 & -160275 & 272 & 0.87 & -1.50 & 0.41 & GSE \\
2538463596221422976 & 18.4185 & 2.1610 & 7.466 & -11.608 & -262 & -167128 & 211 & 0.85 & -1.30 & 0.35 & GSE \\
5783102618581820672 & 192.5538 & -81.9544 & 9.371 & 1.509 & -64 & -140426 & 722 & 0.95 & -1.63 & 0.40 & GSE \\

\hline

\end{tabular} 
\tablefoot{Column (1) {\it Gaia} DR3 \texttt{source\_id}; (2) and (3) Coordinates; (4) and (5) Proper motions; (6) Angular momentum along the $Z$-axis; (7) Total energy; (8) Component of angular momentum perpendicular to $L_z$; (9) Eccentricity; (10) and (11) Photometric metallicities and their corresponding uncertainties; (12) The substructure to which the star is assigned by the CLiMB algorithm.

Columns (1)–(5) are taken from the {\it Gaia} DR3 \texttt{gaia\_source} table \citep{Vallenari2023}.
Columns (6)–(9) and (12) are calculated in this study.
Columns (10) and (11) are from \citet{Muraveva2025}.
The full table is published at the CDS; a portion is shown here to illustrate its form and content.}

\end{table*} 




\end{appendix}

\end{document}